\documentclass{aa}  

\usepackage{graphicx}
\usepackage{subfigure}
\usepackage{subfloat}
\usepackage{float}
\usepackage[version=3]{mhchem}
\usepackage{booktabs}
\usepackage{pdflscape}
\usepackage{supertabular}
\usepackage{longtable,lscape}
\usepackage{txfonts}
\begin{document} 

\title{Determining protoplanetary disk gas masses from CO isotopologues line observations}


\author{A. Miotello\inst{\ref{inst:leiden}}\and
       E. F. van Dishoeck\inst{\ref{inst:leiden},\ref{inst:mpe}}\and M. Kama\inst{\ref{inst:leiden}}\and S. Bruderer \inst{\ref{inst:mpe}}}

\institute{
Leiden Observatory, Leiden University, Niels Bohrweg 2, NL-2333 CA Leiden, The Netherlands\label{inst:leiden}
\and
Max-Planck-institute f{\"u}r extraterrestrische Physik, Giessenbachstra{\ss}e, D-85748 Garching, Germany\label{inst:mpe}
}
                       

\abstract{Despite intensive studies of protoplanetary disks, there is still no reliable way to determine their total (gast+dust) mass and their surface density distribution, quantities that are crucial for describing both the structure and the evolution of disks up to the formation of planets.} 
{The goal of this work is to use less abundant CO isotopologues, such as $^{13}$CO, C$^{18}$O and C$^{17}$O, whose detection is routine for ALMA, to infer the gas mass of disks. Isotope-selective effects need to be taken into account in the analysis, because they can significantly modify CO isotopologues line intensities.} 
{CO isotope-selective photodissociation has been implemented in the physical-chemical code DALI (Dust And LInes) and more than 800 disk models have been run for a range of disk and stellar parameters. Dust and gas temperature structures have been computed self-consistently, together  with a chemical calculation of the main atomic and molecular species. Both disk structure and stellar parameters have been investigated by varying the parameters in the grid of models. Total fluxes have been ray-traced for different CO isotopologues and for various low $J-$ transitions for different inclinations. } 
{A combination of $^{13}$CO and C$^{18}$O total intensities allows inference of the total disk mass, although with non-negligible uncertainties. These can be overcome by employing spatially resolved observations, i.e. the disk's radial extent and inclination. Comparison with parametric models by \cite{Williams14} shows differences at the factor of a few level, especially for extremely low and high disk masses. Finally, total line intensities for different CO isotopologue and for various low-$J$ transitions are provided and are fitted to simple formulae. The effects of a lower gas-phase carbon abundance and different gas-to-dust ratios are investigated as well, and comparison with other tracers is made.} 
{Disk masses can be determined within a factor of a few by comparing CO isotopologue lines observations with the simulated line fluxes provided in this paper, modulo the uncertainties in the volatile elemental abundances.} 


\maketitle
%
\section{Introduction}

Protoplanetary disks are known to be the birth places of planets, but the process of grain growth, and ultimately planetesimal formation depends sensitively on the physical structure of gas in disks \citep[see e.g.,][for a review]{Armitage11}. This structure is still poorly constrained from observations \citep[e.g.,][]{Williams11}. In particular, there is not yet a reliable way for determining the total disk mass (gas + dust) and its radial distribution, which is one of the crucial parameters needed for describing disks' structure and evolution. 

The traditional way of measuring disk masses has been through the detection of the cold dust, which dominates disk emission at millimeter wavelengths and is readily observed \citep[e.g.,][]{Beckwith90,Natta04,Ricci10,Testi14}. The dust component typically accounts for only 1\% of the total disk mass, as the bulk of it is in the molecular gas. A conversion factor of 100 is generally taken for translating from dust to gas mass. However, as recent observations have shown, often the dust distribution does not follow that of the gas \citep[e.g.,][]{Panic09,Andrews12,Casassus13,vanderMarel13,Walsh14,Perez15}. In addition, major assumptions on dust particle size, shape, and composition need to be made in order to convert dust emission into dust mass. Finally, due to disk evolution, the gas-to-dust ratio may deviate significantly from 100 \citep[][]{Gorti09,Alexander14}. In order to overcome these uncertainties, a direct measurement of molecular gas is preferred in order to determine the total gas mass, its spatial distribution and that of the gas-to-dust ratio of protoplanetary disks.

Ideally, one would like to trace directly molecular hydrogen (H$_2$), the most abundant species, whose rotational lines are however difficult to detect. H$_2$ has no permanent dipole moment, requires higher temperatures for excitation and emits at wavelengths where the dust is optically thick. A solution comes from the detection of the less abundant isotopologue HD, whose far-infrared transitions were covered with the \emph{Herschel} Space Observatory. Observations of HD lines have been published only for the closest protoplanetary disk, the well known TW Hya disk \citep{Bergin13}, and no current facility is sensitive enough for detecting HD in disks not covered by \emph{Herschel}. 
These limitations lead to the need of other gas mass tracers. 

The second most abundant molecule, carbon monoxide (CO), is on the other hand readily detectable through its pure rotational lines at millimeter (mm-)wavelengths. However the main isotopologue, $^{12}$CO, becomes optically thick at low column densities and only traces the surfaces of protoplanetary disks. Thus, in order to trace the bulk of the mass, one needs to observe less abundant CO isotopologues like $^{13}$CO and C$^{18}$O, that become optically thick much deeper and trace the gas down to the midplane of the disk \citep{vanZadelhoff01,Dartois03}. Recently \cite{Williams14} have shown that through a combination of observations of multiple CO isotopologues (e.g., $^{13}$CO and C$^{18}$O) with a grid of parametric models it is possible to determine disk gas masses, even if the disk properties are not known. These models do not model chemistry, however, but use a parametrized temperature structure and CO abundance with fixed $\rm[ ^{12}C]/[^{13}C]$ and $\rm[ ^{16}O]/[^{18}O]$ ratios.

Although the CO chemistry is well studied and easily implemented in chemical models of disks \citep[e.g.,][]{Aikawa02,Hollembach05,Gorti09,Nomura09,Woitke09}, this is not the case for isotope-selective processes \citep{Visser09}, which may affect the correct determination of disk masses. Miotello et al. (2014) have implemented isotope-selective photodissociation in a thermo-chemical and physical code, DALI  \citep[Dust And LInes][]{Bruderer12,Bruderer13}, exploring a set of parameters in a small grid of disk models. This study shows that isotope-selective processes may significantly affect CO isotopologues abundances, line emission and, accordingly, the determination of disk masses, especially as inferred from C$^{18}$O and C$^{17}$O alone. 

In this paper we expand on the \cite{Miotello14} study and provide conversion factors to go from CO isotopologue line fluxes to the disk gas mass. A larger grid of disk models, 840 in total, is presented. After running a thermo-chemical calculation including isotope-selective photodissociation, CO isotopologues line intensities have been ray-traced for all models and lines are provided in Appendix \ref{AppendixA} and in electronic tables. These values can be used to infer disk masses, in particular if information about surface density distributions is derived from spatially resolved CO isotopologues observations. We also compare with alternative gas mass tracers such as the atomic fine structure lines \citep[e.g.][]{Kamp11,Dent13,Woitke15}.

\section{Model}
\label{model}
In this work 2D disk models have been run to relate gas mass to CO isotopologues emission.
Our models are computed using the physical-chemical code DALI, extensively tested 
with benchmark test problems \citep[][]{Bruderer12,Bruderer13} and against 
observations \citep[][]{Bruderer12,Fedele13,Bruderer14}.  
The inputs needed by the code are the disk density structure and the stellar spectrum, which provides the central source of heating and dissociating radiation. First, the continuum radiative transfer is solved with a 2D Monte
Carlo method which calculates the dust temperature $T_{\rm dust}$ and 
local continuum radiation field from UV to mm wavelengths. The following step is the determination of the chemical composition of the gas, obtained by a time-dependent chemical network simulation: the abundances of the main atoms and molecules are computed and subsequently the molecular excitation is obtained through a non-LTE calculation. Then the gas temperature $T_{\rm gas}$ is obtained from the 
balance between heating and cooling processes. 
Since both the chemistry 
and the molecular excitation depend on $T_{\rm gas}$, the problem is 
solved iteratively until a self-consistent solution is found. Finally, spectral 
image cubes are created with a raytracer.

As in \cite{Miotello14}, we have included a complete treatment of isotope-selective 
processes. In the chemical network different isotopologues are taken as 
independent species (e.g. $^{12}$CO, $^{13}$CO, C$^{18}$O, and C$^{17}$O) and reactions which regulate the abundance of one isotopolog over the other are included. Mutual (from H, H$_2$, CO) and self-shielding factors for different CO isotopologues from \cite{Visser09} have been implemented in the chemical calculation. 

\subsection{Physical structure}

In our models the disk density structure is parametrized by a power-law function, following the prescription proposed by 
\cite{Andrews11}. Physically this represents a viscously evolving disk, where the viscosity is expressed by $\nu \propto  R^{\gamma}$ \citep{Lynden-BellPringle74,Hartmann98}. The surface density follows:
\begin{equation}
\Sigma_{\rm gas}=\Sigma_c \,\left( \frac{R}{R_c} \right) ^{-\gamma} \exp \, \left[ - \,\left( \frac{R}{R_c} \right) ^{2-\gamma} \right] \ ,
\label{sigma_eq}
\end{equation}
where $R_c$ is the critical radius and $\Sigma_c$ is the critical surface density. A fixed gas-to-dust ratio equal to 100 is employed to set the relation between the gas and the dust surface densities.

The vertical density structure follows a Gaussian with scale height angle 
$h=h_{\rm c} (R/R_{\rm c})^{\psi}$. For the dust settling, two populations of grains are 
considered, small (0.005 - 1 $\mu$m) and large (1 - 1000 $\mu$m) \citep{Dalessio06}. The 
scale height is $h$ for the the small grains and $\chi h$ for the large ones, where 
$0<\chi \leq 1$. 
The fraction of surface density distributed to the small and large population is $(1-f_{\rm large})\Sigma_{\rm dust}$ and $f_{\rm large}\, \Sigma_{\rm dust}$, respectively.
In \cite{Miotello14} the ratio of large to small grains was varied:$f_{\rm large}=0.99$ was chosen in order to simulate a mixture of large and small grains (most of the dust mass in large grains) and $f_{\rm large} = 10^{-2}$ was set for the situation of only small grains (only 1\% of dust mass in large grains). Since small grains are much more efficient in absorbing UV radiation, the output CO isotopologues abundances and lines were different in the two cases. However, no significant differences were found between the case of $f_{\rm large}=0.9$ and of $f_{\rm large}$ between 0.1 and 0.99. Since grains are expected to grow to mm-sizes in protoplanetary disks \citep[e.g.,][]{Testi03,Rodmann06,Lommen09,Ricci10}, for this work we assume only one case, $f_{\rm large}=0.9$.
In contrast with \cite{Miotello14}, the gas vertical density structure is coupled only to the small grains distribution. The  gas-to-dust ratio is enforced on the total column of gas and dust (both small and large grain populations).

Other parameters of the model are the stellar radiation field and the cosmic-ray ionization rate. Particularly important is the FUV (6 -13.6 eV) part of the stellar radiation field. In order to study the effects of different amounts of FUV photons 
compared with the bolometric luminosity, the stellar spectrum is assumed to be a black-body at 
a given temperature $T_{\rm eff}$ with excess UV due to accretion. The X-ray spectrum is taken to be a thermal spectrum of $7 
\times 10^7$ K within 1 - 100 keV and the X-ray luminosity in this band $L_{\rm X} = 10^{30}$ 
erg s$^{-1}$. As discussed in \cite{Bruderer13}, the X-ray luminosity is of minor importance 
for the intensity of CO pure rotational lines which are the focus in this work. The cosmic-ray 
ionization rate is set to $5 \times 10^{-17}$ s$^{-1}$. We account for the interstellar UV 
radiation field and the cosmic microwave background as external sources of radiation.

The calculation is carried out on 75 cells in the radial direction and 60 in the vertical direction. The 
spectral grid of the continuum radiative transfer extends from 912 \AA $\,$ to 3 mm in 58 wavelength-bins. The wavelength dependence of the photo-dissociation and photo-ionization cross sections is taken into account using data summarized in \cite{VD06}.

The chemistry is solved in a time dependent manner, up to a chemical age of 1 Myr. The 
main difference to the steady-state solution is that carbon is not converted into methane 
(CH$_4$) close to the mid-plane, since the time-scale of these reactions is longer than $>10$ 
Myr and thus unlikely to proceed in disks. CO line intensity results are not sensitive to this choice of age for values between 0.5-5 Myr. As in \cite{Miotello14} we run models both with isotope-selective processes switched on (ISO network) or off (NOISO network). The NOISO models do not include isotopologues in their chemical network and for the calculation of line fluxes, e.g. by $^{13}$CO, the abundance is scaled by a constant factor given by the ISM isotope-ratio [$^{13}$C]/[$^{12}$C].

\subsection{Chemical network}
\label{chem_net}

Our chemical reaction network is a reduced version of the ISO network presented by \cite{Miotello14}, which is based on the UMIST 06 network \citep{Woodall07,Bruderer12,Bruderer13}. It includes H$_2$ formation on dust, freeze-out, thermal and non-thermal desorption, hydrogenation of simple species on ices, gas-phase reactions, photodissociation, X-ray and cosmic-ray induced reactions, PAH/small grain charge exchange/hydrogenation, and reactions with vibrationally excited H$_2$. The details of the implementation of these reactions are reported in Appendix A.3.1 of \cite{Bruderer12}, while the implementation of CO isotope-selective processes is discussed in \cite{Miotello14}. Specifically, a binding energy of 855 K was used for pure CO ice \citep{Bisschop06}. The elemental abundances are also the same as in \cite{Bruderer12}. The isotope ratios are taken to be  [$^{12}$C]/[$^{13}$C]=77, [$^{16}$O]/[$^{18}$O]=560 and [$^{16}$O]/[$^{17}$O]=1792 \citep{WilsonRood94}. 

Since CO is the main carbon-bearing species in protoplanetary disks, its abundance is strictly related to that of atomic carbon. As shown in \cite{Bruderer12}, \cite{Favre13}, \cite{Bergin14}, \cite{Du15} and Kama et al. (2015, submitted), some protoplanetary disk observations show evidence for carbon and/or oxygen depletion. So, CO isotopologue line intensities may be fainter because carbon is locked in more complex organics \citep[see e.g.][]{Bergin14}. In order to quantitatively explore the effects of carbon depletion on CO isotopologues line intensities, a small set models has been run with a reduced initial carbon abundance. The standard value assumed for the whole grid,  $X_{\rm C}=\text{C/H}=1.35 \times 10^{-4} $, has been reduced by factors of 3 and 10. Accordingly, the initial atomic abundance of $^{13}$C has been reduced by the same factors. These results and a more detailed discussion about the effects of carbon depletion on CO isotopologues lines are reported in Sec. \ref{C_depletion} and in Appendix \ref{AppendixB}.

In order to be able to run a significantly larger number of models in a feasible amount of time, the ISO chemical network presented in \cite{Miotello14} has been reduced. It was possible to identify which species and reactions channels did not have an impact on CO chemistry, after running some test models with different versions of the chemical network. In particular, sulfur chemistry was found to have negligible influence on the CO chemistry. Accordingly, all sulfur-baring molecules were removed by the list of the accounted species. Sulfur atoms and other metals like Si, Mg and Fe, are kept in the chemical network for allowing charge transfer.
The simplifications applied to the chemical network do not bring significant variations in CO isotopologues abundance distributions and modify their resulting line intensities by less than 1\%.

Finally, the total number of species has been reduced from 276 to 185 and the number of the reactions from 9755 to 5755. Since the time consumed is driven mainly by the number of the species in the chemical network, this reduction allows us to explore a much larger set of parameters, discussed in Sec. \ref{parameters}. From now on, we will refer to this chemical network as the \emph{ISO network}; all models have also been run without accounting for isotope-selective processes, using the \emph{NOISO network} presented in \cite{Miotello14}.

\subsection{Grid of models}
\label{parameters}
The physical structure of protoplanetary disks is difficult to constrain from observations without significant degeneracy in model parameter combination. In particular the vertical structure has been constrained only in few cases directly by observations \cite[e.g.][]{Rosenfeld13}. On the other hand, the structure of a disk may strongly affect the location and the intensity of the molecular emission. We explore to what extent CO isotopologues lines trace $M_{\rm disk}$ by varying the radial and the vertical structure of the disk.
\paragraph{Radial structure.}
In a disk model the surface density distribution sets how the bulk of the gas mass is radially distributed inside the disk. Varying the values of different parameters in Eq. (\ref{sigma_eq}) influences the simulated line intensities. For exploring different scenarios, the following parameters have been varied:
\begin{itemize}
\item[\emph{i.}] The \emph{critical radius} $R_{\rm c}$ has been assumed to be 30 au, 60 au and 200 au, in order to simulate disks with different radial extents. At $R>R_{\rm c}$ the exponential controls the surface density distribution. Indirectly, $R_{\rm c}$ regulates the disk outer radius. \\
\item[\emph{ii.}] The \emph{power-law index} $\gamma$ determines the disk radial size, too. This parameter mainly determines how the mass is distributed along the radial extent of the disk. Shallower power-laws allow the bulk of the mass to be further out in the disk. In this paper $\gamma$ is taken to be 0.8, 1.0, and 1.5. 
\end{itemize}
These ranges of $R_{\rm c}$ and $\gamma$ cover those found observationally by \cite{Andrews12}. Very extended disks with $R_{\rm c}=200$ au are however expected to be extreme cases.
\paragraph{Vertical structure.} From a modeling point of view, different vertical structures can strongly affect the disk temperature structure and the chemistry. Changes in the temperature structure affect the molecular composition and thus alter the emitting CO column density.
\begin{itemize}
\item[\emph{i.}] The \emph{scale height} $h_{\rm c}$, defined by $h=h_{\rm c} (R/R_{\rm c})^{\psi}$, regulates the vertical extent of the disk. In this grid of models $h_{\rm c}$ is allowed to be either 0.1 or 0.2.\\
\item[\emph{ii.}] The \emph{flaring angle} $\psi$ indicates how much the disk spreads in the vertical direction and thus how much direct stellar radiation it is able to intercept. Also $\psi$ has been taken to have two values, 0.1 and 0.2 rad. 
\end{itemize}
\paragraph{Mass and radiation field.} Finally, the total disk mass is the main parameter governing CO isotopologues column densities and line intensities. At the same time, the amount of photodissociating photons plays an important role for the survival of CO isotopologues. In order to explore different combinations of column densities and FUV fluxes, disk masses and stellar spectra have been varied.
\begin{itemize}
\item[\emph{i.}] The total \emph{disk mass} has been covered for seven values in a realistic range: $M_{\rm disk}=10^{-5}, 10^{-4}, 5\times10^{-4}, 10^{-3}, 5\times10^{-3}, 10^{-2}, 10^{-1} M_{\odot}$.\\
\item[\emph{ii.}] The stellar flux has been assumed to be a black body with an effective temperature $T_{\rm eff}=4000$ K or $T_{\rm eff}=10000$ K to simulate either a T Tauri or a Herbig star, as in \cite{Miotello14}. Excess UV radiation due to accretion is taken into account for T Tauri stars as follows.The gravitational potential energy of accreted mass is assumed to be released with 100$\%$ efficiency as blackbody emission with $T$=10000 K. The mass accretion rate is taken to be 10$^{-8} M_{\odot}$ yr$^{-1}$ and the luminosity is assumed to be emitted uniformly over the stellar surface. An accretion rate of $10^{-8} M_{\odot} \rm yr^{-1}$ is the typical value measured in classical Class II disks \citep{Hartmann98,Mohanty05,Herczeg_Hillenbrand08,Alcala14,Manara16} and lack of significant accretion variability was found for these kinds of objects \citep{Costigan14}. This results in $L_{\rm FUV}/L_{\rm bol}$= 1.5$\cdot10^{-2}$ for the T Tauri and  $L_{\rm FUV}/L_{\rm bol}$= 7.8$\cdot10^{-2}$ for the Herbig stars. The interstellar radiation field is also included.
\end{itemize}

Finally, two values for the \emph{inclination angle} $i$ have been chosen for the ray-tracing, 10$^{\circ}$ and 80$^{\circ}$. Line luminosities are not expected to vary significantly for intermediate inclinations, i.e., up to 70$^{\circ}$ \citep{Beckwith90}. The results obtained for $i = 10^{\circ}$ can be considered for the analysis of typical disks. On the other hand, line intensities are extremely reduced in the rare case of an edge-on disk, i.e., with an inclination close to 90$^{\circ}$. In order to account for such objects, disk models with $i = 80^{\circ}$ have been run. For the ray-tracing all the modeled disks are assumed to be at a distance of 100 pc from Earth.

All parameters explored with this grid of models are summarized in
Table \ref{tab:modelpar}. The gas-to-dust ratio is generally kept fixed to 100
because it is not in the scope of this paper to compare dust-based and
gas-based disk mass determinations. However, this parameter can and
should be set as variable when interpreting combined continuum and line
observations and some initial exploration of the effects of
  lowering the gas-to-dust ratio are presented below. Note that one
  can study these effects by either keeping the gas mass or
  the dust mass constant. Here the former approach is taken:
  keeping the gas mass fixed. Other studies have explored the
  sensitivities of line intensities by keeping the dust mass fixed and
  changing the gas mass \citep[e.g.,][]{Bruderer12,Kama16,Woitke15}.

\begin{table}[tbh]
\caption{Parameters of the disk models.}
\label{tab:modelpar}
\centering
\begin{tabular}{ll}
\hline\hline
Parameter	 & Range \\
\hline
\emph{Chemistry} \\
Chemical network & ISO / NOISO \\
Chemical age & 1 Myr \\
$\rm [C]/[H]$ & 1.35$\times 10^{-4}$\\
$\rm [PAH]$ & $10^{-2}$ ISM abundance \\
\emph{Physical structure}& \\
$\gamma$ & 0.8, 1, 1.5\\
$\psi$ & 0.1, 0.2 \\
$h_{\rm c}$ & 0.1, 0.2 rad \\
$R_{\rm c}$ & 30, 60, 200 au \\
$M_{\rm gas}$ & $10^{-5} , 10^{-4} , 5\times10^{-4}, 10^{-3},$ \\
&  $5\times10^{-3},  10^{-2} , 10^{-1} M_\odot$\\
Gas-to-dust ratio & 100, 10, 1\\
$f_{\rm large}$ & 0.9\\
$\chi$ & 1 \\
$i$ & $10^{\circ},\, 80^{\circ}$ \\
\emph{Stellar spectrum}&\\
$T_{\rm eff}$ & 4000 K +Accretion UV,\\
&10000 K\\
$L_{\rm bol}$ & 1, 10 $L_{\odot}$\\
$L_{\rm X}$ & $\rm 10^{30}\, erg\,s^{-1}$\\
\emph{Dust properties}&\\
Dust & 0.005-1 $\mu$m (small)\\
& 1-1000 $\mu$m (large)\\
\hline
\end{tabular}
\end{table}

\section{Results}
\label{results}
\begin{figure}
   \resizebox{\hsize}{!}
             {\includegraphics[width=1.\textwidth]{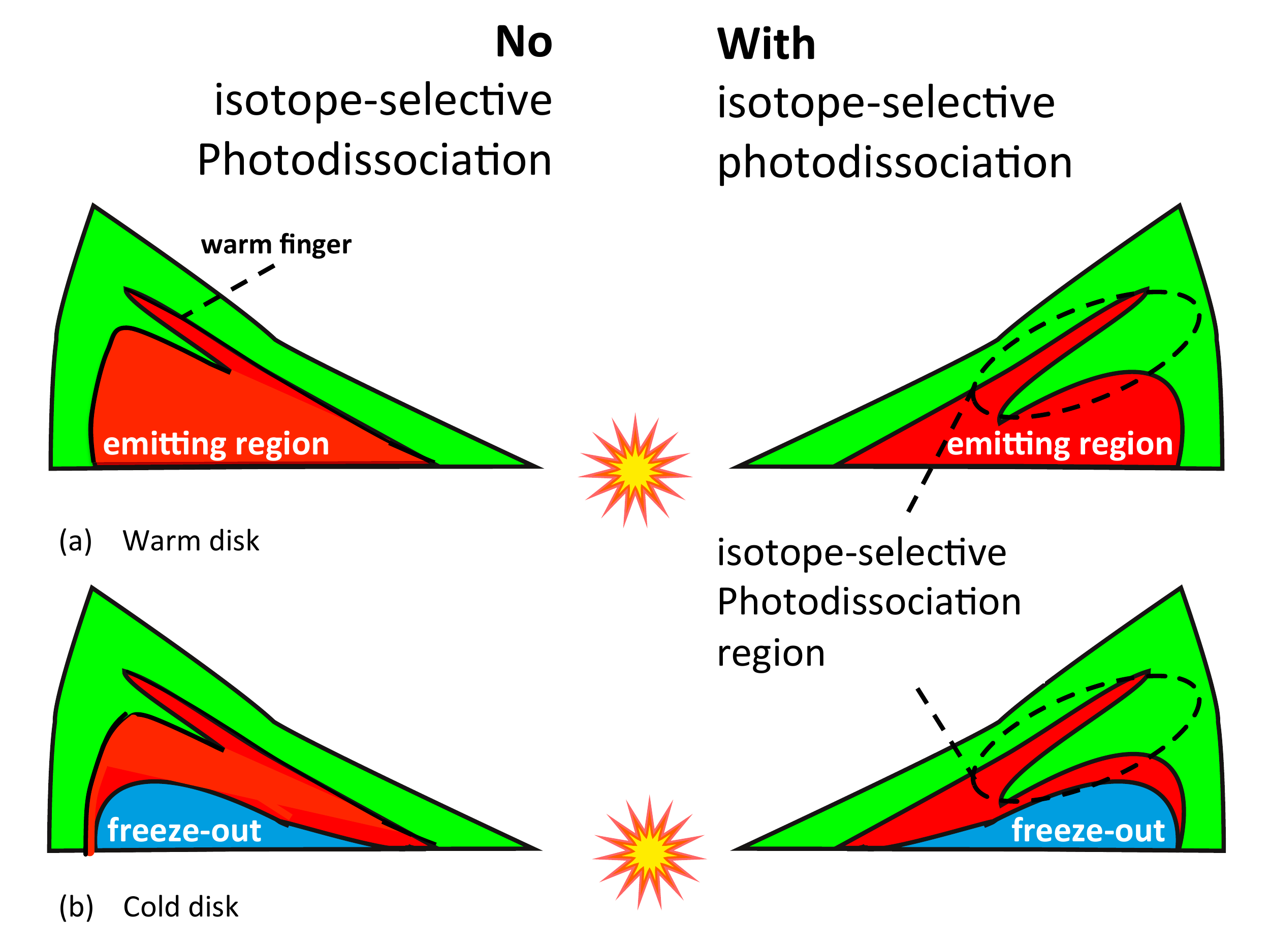}}
      \caption{Sketch of a section of a disk. On the left hand side the region in red shows where CO isotopologues are expected to survive, if isotope-selective dissociation is not included. On the right hand side isotope-selective effects are included and the emitting region in red is reduced. Panel (a) shows the case of a warm disk, where CO isotopologues do not freeze-out anywhere in the disk, while panel (b) shows the case of a colder disk where freeze-out occurs in the midplane. The combination of freeze-out and isotope-selective photodissociation (panel (b), right hand side) drastically reduces CO isotopologues emission.}
       \label{sketch}
\end{figure}

\subsection{Abundances}
Photodissociation is the main process that regulates the abundance of gas phase CO in the emitting layer of disks and it is also the main isotope-selective process \citep[see][and references therein]{Visser09}. The UV absorption lines, that initiate the dissociation process, can become optically thick and CO can shield itself from photodissociating radiation. For the main isotopologue, $^{12}$CO, UV absorption lines become optically thick already at the surface of the molecular layer in disks. On the other hand, less abundant isotopologues (e.g., $^{13}$CO, C$^{18}$O and C$^{17}$O) shield themselves at higher column densities, i.e. deeper in the disk and closer to the midplane. As shown in \cite{Miotello14}, this results in a region where $^{12}$CO is self-shielded and able to survive, while less abundant isotopologues are still photodissociated. In these regions the rarer isotopologues (e.g.,  C$^{18}$O) are underabundant when compared to the correspondent ISM isotopic ratio ([$^{18}$O]/[$^{16}$O]), that is usually adopted for predicting CO isotopologue abundances. 

Fig. \ref{sketch} shows a sketch of a disk section. On the left hand side the red regions show where CO isotopologues survive if isotope-selective dissociation is not included. On the right hand side isotope-selective effects are included and the emitting region in red is reduced. Panel (a) shows the case of a warm disk, where CO isotopologues do not freeze-out anywhere in the disk, while panel (b) shows the case of a colder disk where freeze-out occurs in the midplane. The combination of freeze-out and isotope-selective photodissociation (panel (b), right hand side) drastically reduces CO isotopologues column densities. Warmer disks, generally found around Herbig stars, are however also exposed to stronger UV fluxes compared with T Tauri disks. This causes photodissociation to be more efficient in Herbig disks and, accordingly, CO isotopologue abundances are strongly decreased. A more quantitative discussion on the competition between temperature effects and UV irradiation is presented in Sec. \ref{stell_par}.

Comparison of models with the same disk parameters but run alternatively with the ISO and the NOISO network show that isotope-selective photodissociation has more effects on C$^{18}$O and C$^{17}$O abundances, than on $^{13}$CO survival. C$^{18}$O and C$^{17}$O present regions of significant underabundance (more than one order of magnitude) if isotope-selected processes are implemented in the models, while $^{13}$CO shows a slight overabundance of a factor of two \citep[see Figs. 3 and 6 in][]{Miotello14}. C$^{18}$O and C$^{17}$O behave in an almost identical manner, therefore C$^{17}$O results are omitted in the following discussion. On the other hand, C$^{17}$O integrated line intensities are reported in Table \ref{tab_lineint}.

\begin{figure*}
   \resizebox{\hsize}{!}
             {\includegraphics[width=0.8\textwidth]{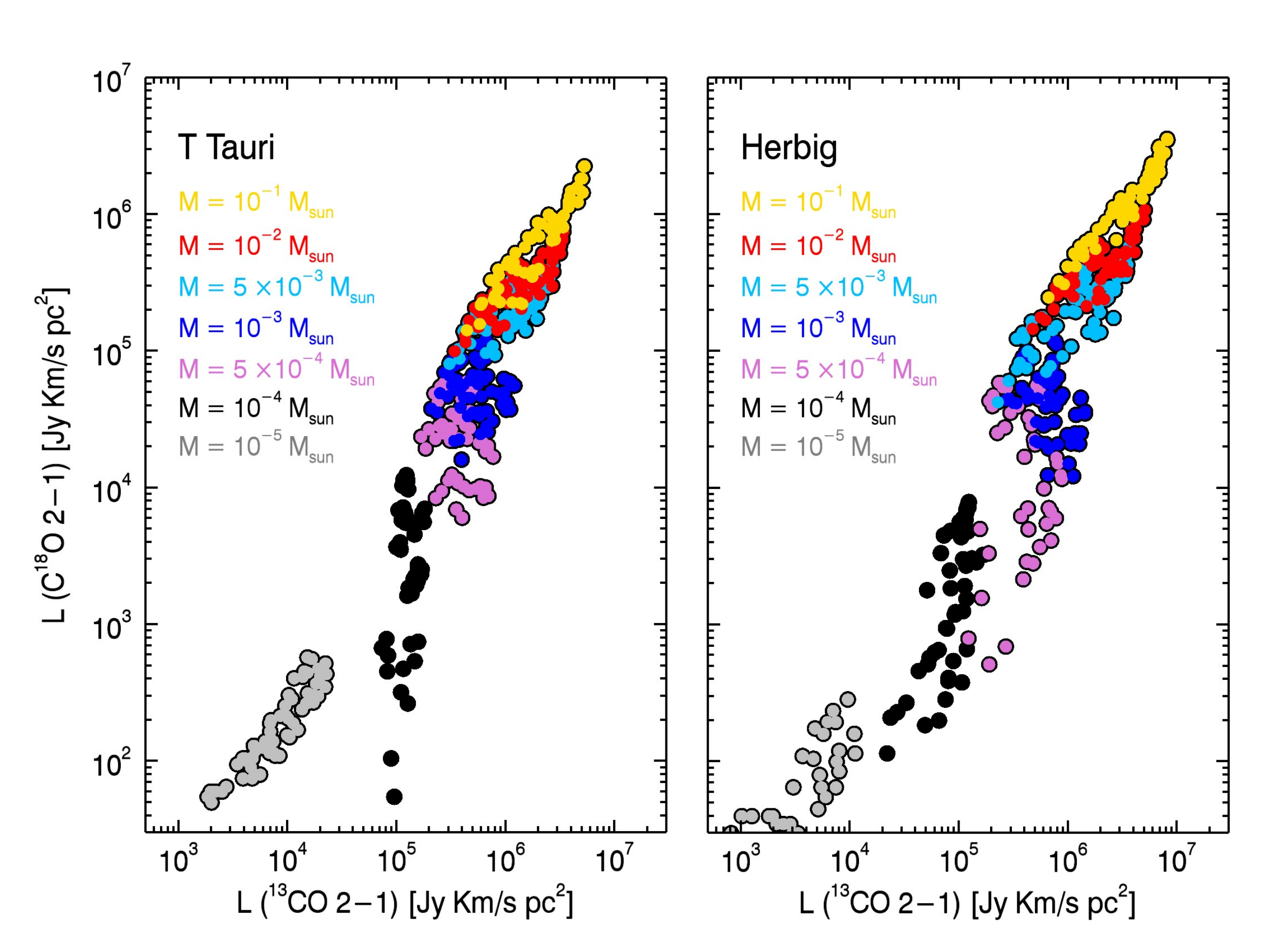}}
      \caption{C$^{18}$O vs $^{13}$CO ($J=2-1$) line luminosity obtained implementing isotope-selective processes. Different colors represent models with different disk gas masses. The scatter of the points is due to the exploration of various disk parameters ($\gamma, \, R_{\rm c}, \psi, h_{\rm c},\, \text{and } i$).  Models shown in the left panel represent T Tauri disks , while those in the right panel represent Herbig disks, all located at a distance of 100 pc.}
       \label{clouds}
\label{lines}
\end{figure*}

\subsection{Line intensities}
\label{line_int}

\subsubsection{Line intensities and disk masses}

Our model results show that a combination of disk spatially integrated $^{13}$CO and C$^{18}$O observations allow to assign a value for the total disk gas mass for a given carbon abundance. This result is similar to that found in \cite{Williams14}, with important differences described in Sec. \ref{comparison}. Fig. \ref{clouds} presents the $J=2-1$ transitions of $^{13}$CO vs C$^{18}$O for the entire grid of models with colors indicating the different disk masses. It is seen that the points in this diagram show a clear disk mass segregation of the results, no matter which parameters are assumed in the disk models. This is particularly true for low-mass disks, $M_{\rm disk}=10^{-5} - 10^{-3}M_{\odot}$, while in the higher mass regime the separation becomes less clear. Here disk mass determinations are degenerate up to two orders of magnitude. This is caused by the high level of CO freeze-out in more massive disks. Only a small fraction of CO is left in the gas-phase in the outer disk and increasing the disk mass does not add much emission, defining an upper threshold  for CO isotopologues line intensities. Accordingly, warmer Herbig disks (right panel of Fig. \ref{clouds}) present a lower level of freeze-out and reach higher line luminosities. In Fig. \ref{clouds} results for all models are reported, while in the following plots some of them (those with $M_{\rm disk}=5\times 10^{-4} \text{ and } 5\times 10^{-3} M_{\odot}$) are removed to better visualize the trends. Also, for all disk models transitions of different CO isotopologues up to $J=6-5$ are ray-traced for different disk inclinations, and are reported in Appendix \ref{AppendixA}.

The trends are similar for higher $J$ transitions. As shown in Fig. \ref{56}, the scatter is however enhanced compared with that found for $J=2-1$ line intensities (Fig. \ref{clouds}). Higher $J$ transitions, such as $J=6-5$, indeed trace warmer gas in upper disk layers that are more sensitive to disk structure, while lower transitions are excited where most of the gas disk mass lies. Accordingly, lower $J$ transitions of less abundant CO isotopologues are better mass tracers.

\begin{figure*}
\centering
             {\includegraphics[width=0.7\textwidth]{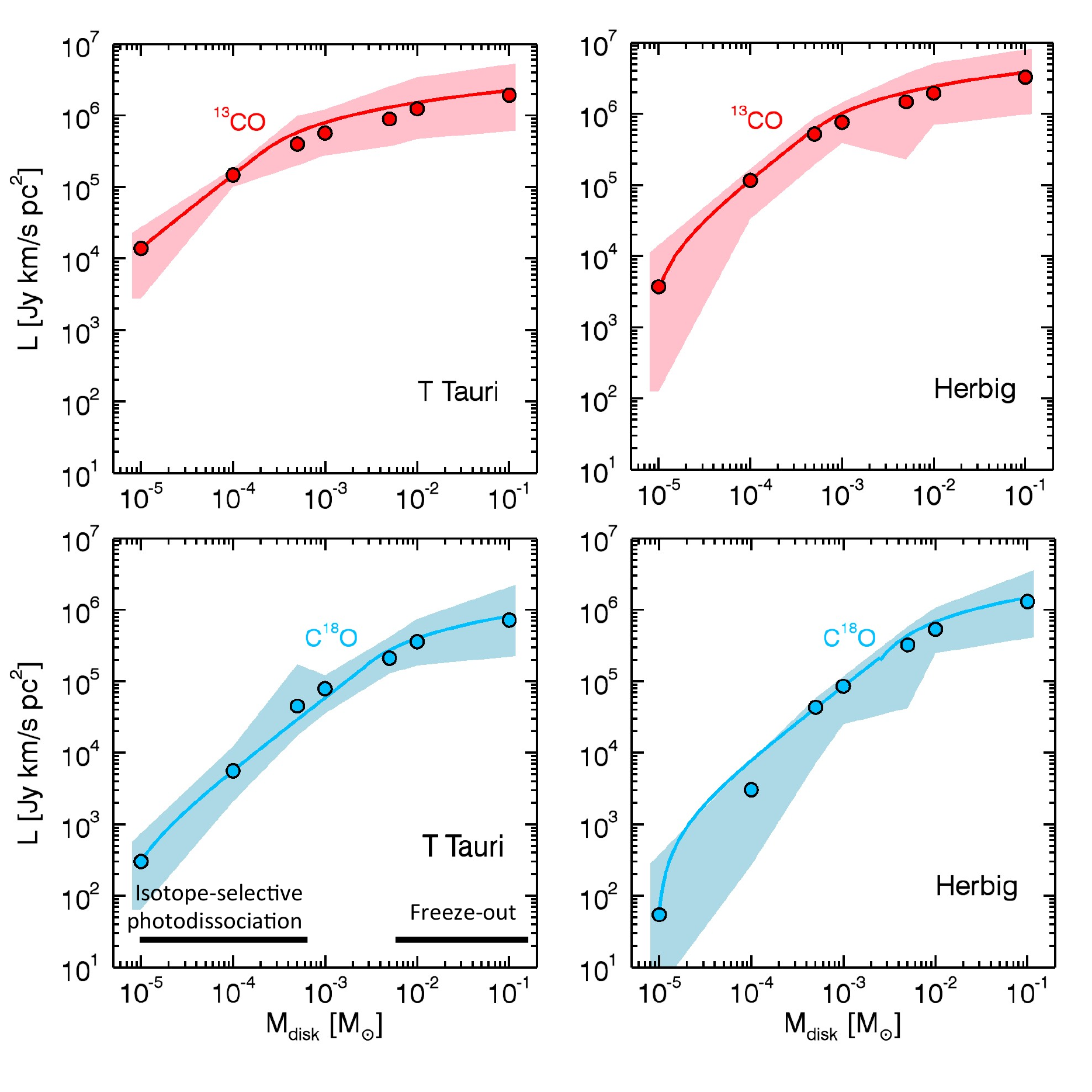}}
      \caption{Median of the $^{13}$CO (in red) and C$^{18}$O (in blue) $J=2-1$ line intensities in different mass bins for models with $i=10^{\circ}$ are presented with filled circles. Solid lines show the fit functions presented in Eq. \ref{L_13}. Results for T Tauri disks are shown in the left panel, while those for Herbig disks in the right panel. The filled regions show the maximum range of the line luminosities obtained with different models.}
       \label{L_mass}
\end{figure*}

It is possible to determine how line intensities depend on the disk mass by computing the medians of the $^{13}$CO and C$^{18}$O $J=2-1$ line intensities obtained by models in different disk mass bins. These trends are presented in Fig. \ref{L_mass}, both for T Tauri and Herbig disks. For high-mass disks ($M_{\rm disk} > 10^{-3}$), line intensities do not increase linearly with mass, but tend to converge to a constant value. As discussed above, this is caused by the high level of CO freeze-out in more massive disks. Accordingly, this convergent trend occurs earlier (i.e., at lower disk masses) for colder T Tauri disks, than for warmer Herbig disks.

The median values of the line luminosities for models with $i=10^{\circ}$ are reported in Table \ref{tab_median} together with their standard deviations. The $^{13}$CO and C$^{18}$O $J=2-1$ line luminosities can be expressed by fit functions of the disk mass:
\begin{equation}
L_{y} =
  \begin{cases}
A_y+B_y\cdot M_{\rm disk} & \quad M_{\rm disk} \leq M_{\rm tr}\\
C_y + D_y\cdot  \text{log}_{10}(M_{\rm disk}) & \quad M_{\rm disk} > M_{\rm tr},\\
  \end{cases}
\label{L_13}
\end{equation}

where $y=13$ or 18, for $^{13}$CO and C$^{18}$O respectively. For low mass disks the line luminosity has a linear dependence on the disk mass, while for more massive disks the trend is logarithmic. The transition point $M_{\rm tr}$ is different for the two isotopologues beacuse $^{13}$CO becomes optically thick at lower column densities than C$^{18}$O. 
The polynomial coefficients $A_y$, $B_y$, $C_y$, and $D_y$, as well as the transition masses $M_{\rm tr}$ are reported in Table \ref{A_mass}. 

Finally it is worth investigating how the implementation of isotope-selective effects changes the obtained line intensities. Results for both T Tauri and Herbig disk models obtained with the NOISO network are reported in Sect. \ref{comparison}.  
The ratios NOISO/ISO of the $^{13}$CO and C$^{18}$O median $J=2-1$ line luminosities for each disk mass bin can be used to quantify this behavior. The ratios are presented in Table \ref{tab_median} together with the standard deviations of the line luminosities for different mass bins. It is seen that if isotope selective effects are not implemented in the models C$^{18}$O lines are overestimated, in particular in the low mass regime. A similar behavior is found for the less abundant isotopologue C$^{17}$O. On the other hand, $^{13}$CO line intensities are slightly underestimated by the NOISO network, except for very low mass disks ($M_{\rm disk}=10^{-5} M_{\odot}$) where they are slightly overestimated. Finally, it is found that isotope-selective effects are more effective for Herbig than for T Tauri disks, in particular in the low mass regime (see fifth column of Table \ref{tab_median}).

\begin{table}[h]
\caption{The first and the third columns report median CO isotopologues line luminosities, $L_{\rm I}$ in Jy km/s pc$^2$, obtained implementing isotope-selective photodissociation (ISO network) together with their standard deviations. The second and the fourth columns list the medians of luminosity ratios, obtained by dividing the results from the simple NOISO and ISO networks. Results for both T Tauri and Herbig disks are reported.}
\label{tab_median}
\centering
\begin{tabular}{lcccc}
\toprule
$M_{\rm disk} [M_{\odot}]$&\multicolumn{2}{c}{$^{13}$CO (2-1)}&\multicolumn{2}{c}{C$^{18}$O (2-1)}\\
\cmidrule(lr){2-3}
\cmidrule(lr){4-5}
&$L_{\rm I}$&$L_{\rm N}/L_{\rm I}$&$L_{\rm I}$&$L_{\rm N}/L_{\rm I}$\\
\midrule
&\multicolumn{4}{c}{T Tauri}\\
\cmidrule(lr){2-5}
$10^{-5}$&$1.4\pm0.5 \cdot 10^{4}$&1.4&$3.0\pm1.3 \cdot 10^{2}$&14\\
$10^{-4}$&$1.5\pm0.4 \cdot 10^{5}$&0.7&$5.6\pm2.3 \cdot 10^{3}$&4.3\\
$10^{-3}$&$5.6\pm2.6 \cdot 10^{5}$&0.4&$7.5\pm2.9 \cdot 10^{4}$&0.9\\
$10^{-2}$&$1.2\pm0.7 \cdot 10^{6}$&0.4&$3.6\pm1.9 \cdot 10^{5}$&0.9\\
$10^{-1}$&$1.9\pm1.0 \cdot 10^{6}$&0.4&$7.2\pm3.8 \cdot 10^{5}$&0.4\\
\hline
&\multicolumn{4}{c}{Herbig}\\
\cmidrule(lr){2-5}
$10^{-5}$&$3.7\pm3.4 \cdot 10^{3}$&3.3&$5.4\pm4.1\cdot 10^{1}$&29\\
$10^{-4}$&$1.2\pm0.5 \cdot 10^{5}$&1.4&$3.0\pm2.7 \cdot 10^{3}$&18\\
$10^{-3}$&$7.6\pm2.8 \cdot 10^{5}$&0.7&$8.5\pm2.6 \cdot 10^{4}$&2.3\\
$10^{-2}$&$1.9\pm0.6 \cdot 10^{6}$&0.5&$5.3\pm2.1 \cdot 10^{5}$&0.6\\
$10^{-1}$&$3.3\pm1.7 \cdot 10^{6}$&0.5&$1.3\pm0.4 \cdot 10^{6}$&0.5\\
\hline
\end{tabular}
\end{table}

\subsubsection{Line ratios}
\begin{figure*}
\centering
             {\includegraphics[width=0.7\textwidth]{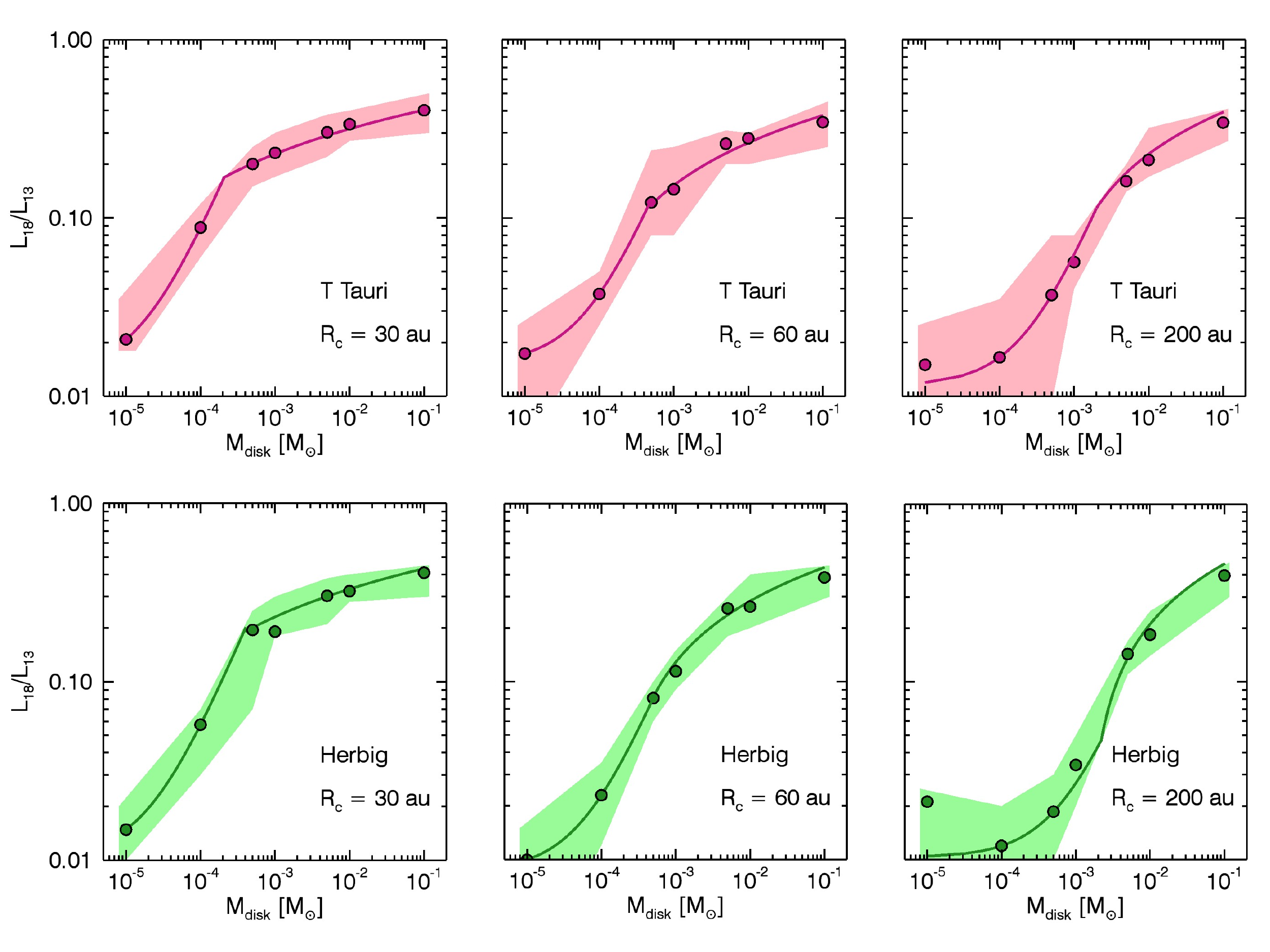}}
      \caption{C$^{18}$O/$^{13}$CO ($J=2-1$) line ratios, defined as $L_{18}/L_{13}$, as function of disk masses for models where $i=10^{\circ}$. The top panels report the results for T Tauri disk models and lower panels those for Herbig disk modelsl. Left, middle and right panels present results for models with respectively $R_{\rm c}= 30$ au, $R_{\rm c}= 60$ au, and $R_{\rm c}= 200$ au. Each point reports the result for the median line ratio in each mass bin.  The continuous solid lines show the fit function reported in eq. (\ref{polin}). The filled regions show the standard deviation from the median values.}
       \label{trend_ratios}
\end{figure*}

An alternative way to present the results is to plot  the ratios of C$^{18}$O/$^{13}$CO $J=2-1$ line intensities, $L_{18}/L_{13}$, as functions of the disk mass (see Fig. \ref{trend_ratios}).
Line ratios obtained by disk models with same $R_{\rm c}$ show a clear trend with disk mass, as shown in Fig. \ref{trend_ratios}. Therefore, if $R_{\rm c}$ can be determined by spatially resolved observations these trends can be employed for determining the disk gas mass with lower uncertainties.  More precisely, as done for line luminosities, the line ratios $L_{18}/L_{13}$ for models with $i=10^{\circ}$ can be fitted by functions of the disk mass:

\begin{equation}
L_{18}/L_{13} =
  \begin{cases}
A_{R_{\rm c}}+B_{R_{\rm c}}\cdot M_{\rm disk} & \quad M_{\rm disk} \leq M_{\rm tr}\\
C_{R_{\rm c}} + D_{R_{\rm c}}\cdot \text{log}_{10}(M_{\rm disk}) & \quad M_{\rm disk} > M_{\rm tr}.\\
  \end{cases}
\label{polin}
\end{equation}

The coefficients $A_{R_{\rm c}}$, $B_{R_{\rm c}}$, $C_{R_{\rm c}}$, and $D_{R_{\rm c}}$ vary for different values of $R_{\rm c}$ and are listed in Table \ref{poly_coeff}. Similarly to eq. \ref{L_13}, the transition from the linear to the logarithmic dependence of the line ratios on the disk mass varies for different values of $R_{\rm c}$. The values for the transition masses $M_{\rm tr}$ are reported in Table \ref{poly_coeff}. If $R_{\rm c}$ is known, the degeneracy in disk mass determinations, shown by the shaded regions in Fig. \ref{L_mass}, is reduced by a factor between 2 and 5.

\subsubsection{Dependence on disk and stellar parameters}
\label{stell_par}

\begin{figure}
   \resizebox{\hsize}{!}
             {\includegraphics[width=1.\textwidth]{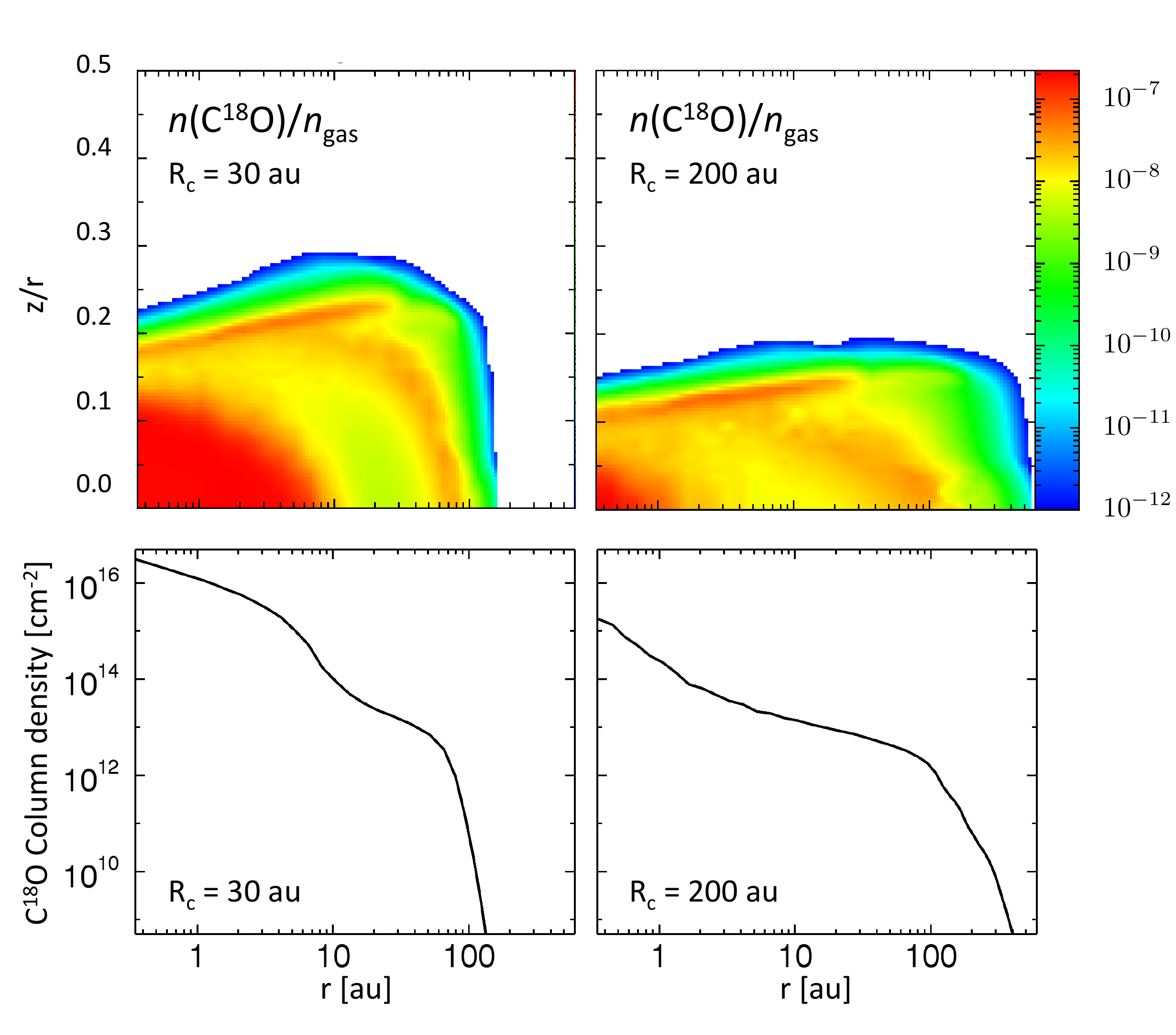}}
      \caption{C$^{18}$O abundances (top panels) and column densities (lower panels) for models with different critical radii: $R_{\rm c}=30$ au in the left panels and $R_{\rm c}=200$ au in the right panels. For this particular T Tauri models the disk parameters are fixed to: $M_{\rm disk}=10^{-5}M_{\odot}, \, \gamma=0.8, \, h_{\rm c}=0.1, \text{ and } \psi=0.1$.}
       \label{Rc}
\end{figure}

It is interesting to investigate how CO isotopologues line intensities depend on some disk parameters (see Fig. \ref{param}). From an observational point of view the radial disk extent can be constrained with spatially resolved data. In this context the first parameter to check is the critical radius $R_{\rm c}$. Line intensities for models with different values of $R_{\rm c}$ are compared in the upper panels of Fig. \ref{param} for the case of T Tauri disks. Very extended disks ($R_{\rm c}=200$ au, see panel c) have lower C$^{18}$O and $^{13}$CO line intensities than more compact disks ($R_{\rm c}=30$ au, see panel a) in the low mass regime. Keeping $M_{\rm disk}$ unchanged, more extended disks have lower column densities at each radius and FUV photons can penetrate easier. As consequence the gas temperature is higher, the photodissociating radiation field is stronger and CO isotopologues are more efficiently dissociated and less abundant. These effects are illustrated in Fig. \ref{Rc}, which shows C$^{18}$O abundances and column densities as functions of radius for two disks with $R_{\rm c}=30$ and 200 au, respectively.

On the other hand, in the high mass regime ($M_{\rm disk}=10^{-1}M_{\odot}$) more extended disks show higher line luminosities than more compact disks. Given the high column densities, CO isotopologues can shield themselves and survive to high abundances at larger radii, if the disk is more extended. The bulk of the emission comes from the outer regions. This holds in particular for $^{13}$CO line emission which is optically thick.

A second parameter to explore is $\gamma$, the power-law index of the surface density distribution. Models with different $\gamma$ values are presented separately in the lower panels of Fig. \ref{param}. The line luminosities produced by models with $\gamma=0.8$ and $\gamma=1$ are almost indistinguishable, while differences can be found comparing these with models with $\gamma=1.5$. In particular in the intermediate disk mass regime, models for $\gamma=0.8,1$ show three different populations that reflect models with different $R_{\rm c}$ values. This effect is less clear for models with $\gamma=1.5$.

Finally, it is interesting to investigate how much the central star influences the CO isotopologues emission in the disk. Models with the same physical disk parameters, but irradiated by a T Tauri-like and a Herbig-like star are presented in Fig. \ref{clouds} (respectively in the left and in the right panel) and in Fig. \ref{L_mass}. For the Herbig disks, the derived line intensities are higher for more massive disks and lower for less massive disks compared with those obtained for the T Tauri disks. Since Herbig disks are warmer, less material is frozen-out onto grains in the higher mass regime and can contribute to the total emission enhancing the line luminosity. Also, the higher temperatures enhance the moderately optically thick $^{13}$CO lines. On the other hand, the strong UV flux coming from the Herbig star is very efficient in photodissociating CO isotopologue molecules for lower mass disks ($M_{\rm disk}=10^{-4}, 10^{-5} M_{\odot}$) where the column densities are not high enough for shielding the photons. This combination of high UV field and low column densities reduces the line luminosities, thus they are lower than those from T Tauri disks.

\subsubsection{Dependence on gas-to-dust ratio}
\label{gas-dust}

To test the sensitivity of the results to the gas-to-dust
  ratio for a fixed gas mass, two T Tauri disk models ($R_{\rm c}=60$
  au, $\gamma=1$, $\psi=0.2$, $h_{\rm c}=0.2$,
  $M_{\rm disk}=10^{-4}, 10^{-2}M_{\odot}$) have been run with reduced
  gas-to-dust ratios of 10 and 1. The ray-traced $J=3-2$ line
  intensities for CO isotopologues are presented in
  Fig.~\ref{check_g_d} as functions of the gas-to-dust ratio,
  normalized to the value at gas-to-dust=100. For the
  $10^{-4}M_{\odot}$ disk the optically thin C$^{18}$O intensities
  increase by a factor of 3.5 if the gas-to-dust ratio is lowered by
  a factor of 100.  This can be understood by the fact that if the
  dust mass is enhanced for a fixed gas mass, the FUV flux will be
  more efficiently attenuated. Accordingly CO and its isotopologues
  survive more easily and their lines are brighter for low gas-to-dust
  ratios, especially if their emission is optically thin. In addition,
  if the dust opacity dominates over the line mutual shielding,
  isotope-selective processes are minimized.  The $^{13}$CO and
  $^{12}$CO emission on the other hand stay the same within 30\%
  because they are more optically thick. There are also small temperature differences between the models of order 20\% in the emitting layer which can affect their line intensities. 

For the $10^{-2}M_{\odot}$ disk the trend is opposite: line
  intensities decrease if the gas-to-dust ratio is lowered for all
  three isotopologues. This happens because the dust column densities
  are increased to the point that the dust becomes optically thick to
  CO isotopologues emission. Therefore reducing the gas-to-dust ratio
  for large disk masses, i.e., adding more dust, results in a larger
  fraction of gaseous CO not contributing to the line emission. 

Overall, it is important to note that the variations in the CO
  isotopologues line fluxes are small for these two test cases, i.e.,
  a factor of a few or less over two orders of magnitude in
  gas-to-dust ratios, keeping the gas mass fixed. This variation is
  much smaller than if the dust mass is kept fixed but the gas mass is
  changed \citep{Bruderer12,Kama16,Woitke15}. Thus, the CO isotopologue line
  fluxes remain a robust tracer of the gas mass, even for low
  gas-to-dust ratios.

\subsubsection{Spatially resolved lines}
\label{sp_res}
Spatially integrated CO isotopologues line intensities are needed for determining the total disk gas content, with special attention to not losing the weak emission coming from the outer disk. However, as discussed above, disk mass determinations are degenerate since they depend on various parameters. Spatially resolved lines may help in breaking this degeneracy, constraining some of the disk parameters usually fixed by the models' inputs. 

\begin{figure*}
   \resizebox{\hsize}{!}
             {\includegraphics[width=1.\textwidth]{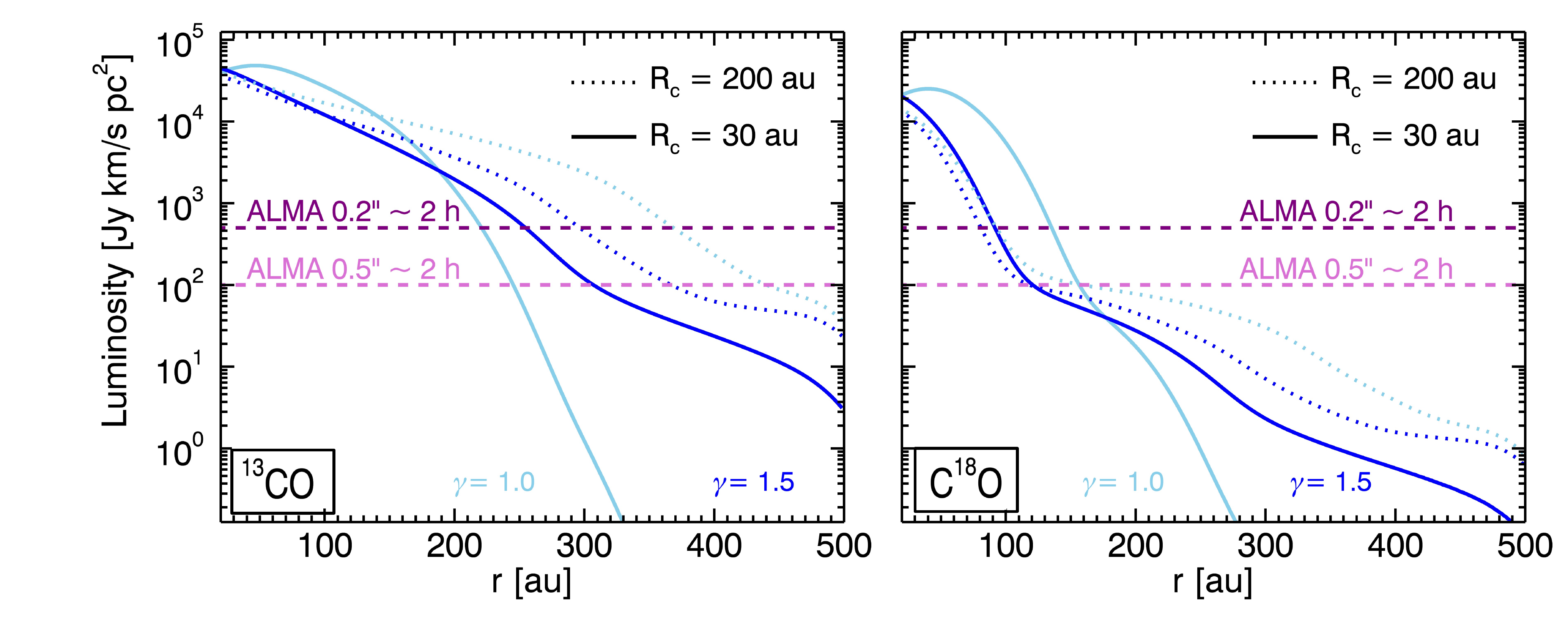}}
      \caption{Radial profile of $^{13}$CO and C$^{18}$O $J=2-1$ lines for four T Tauri models where $\psi=0.1$, $h_{\rm c}=0.1$, $M_{\rm disk}=10^{-3} M_{\odot}$ and $R_{\rm c}$ and $\gamma$ assume different values. Models with $\gamma=1.5$ are presented in dark blue, while those with $\gamma=1.0$ are shown in light blue. Solid lines displays models with $R_{\rm c}=30$ au and dotted lines represents those with $R_{\rm c}=200$ au. These line profiles are obtained assuming a distance of 100 pc, an inclination angle $i=10^{\circ}$, and by convolving the output images by a 0.5" beam. The horizontal dashed lines shows the typical ALMA sensitivity for an integration time of 1.84 hours on source for a 0.2" beam (purple line) and for a 0.5" beam (pink line), given a spectral resolution of 1.3 km/s.}
       \label{conv}
\end{figure*}

The disk's spatial extent is the most straightforward characteristic that spatially resolved observations can constrain. This allows determination of $R_{\rm c}$ and $\gamma$. From an observational point of view, this is done by plotting spatially resolved optically thin line intensities as a function of the disk radial extent (i.e., distance from the central star). In this context, C$^{18}$O and $^{13}$CO are good tracers of the surface density, being optically thin and detectable by ALMA. For example, Fig. \ref{conv} shows the radial profile of $^{13}$CO and C$^{18}$O $J=2-1$ lines for four models where $\psi=0.1$, $h_{\rm c}=0.1$, $M_{\rm disk}=10^{-3} M_{\odot}$ and $R_{\rm c}$ and $\gamma$ assume different values. The modeled line intensities have been convolved with a 0.5" beam in order to simulate typical moderate angular resolution ALMA observations. It is seen that the maximum radial extent of the $^{13}$CO emission is clearly different if $R_{\rm c}=30$ au or if $R_{\rm c}=200$ au, although deep integrations are needed

Furthermore, the power-law spectral index of the surface density distribution may be constrained by spatially resolved observations. As shown by the purple line in Fig. \ref{conv}, with an integration of around 2 hours and an angular resolution of 0.2" it is possible to discriminate between a power-law index $\gamma=1.0$ and $\gamma=1.5$ with $^{13}$CO line emission. 

\section{Discussion}
\label{discussion}

\subsection{Comparison with parametric models}
\label{comparison}
A large grid of parametric models has been run by \cite{Williams14} in order to simulate a CO isotopologues survey for a sample of protoplanetary disks. The work presented here has been partially inspired by this paper and designed to have results that could be easily compared with those by \cite{Williams14}.

The main differences between the two approaches are in how the gas temperature structure is defined and how the CO isotopologues survival is determined. \cite{Williams14} do not calculate the temperature structure and do not run any chemical model. The gas temperature profile is purely parametric, but guided by theory and observations and is a combination of midplane and disk atmosphere temperatures, both assumed to be radial power-laws. This temperature parametrization avoids any assumption about stellar properties and interstellar irradiation. In order to account both for CO freeze-out and photodissociation, gas phase CO is abundant at a constant (ISM-like) abundance for temperatures higher than 20 K and for column densities higher than $1.3 \times 10^{21} \rm H_2 \, cm^{-2}$, as inspired by \cite{vanZadelhoff01,Visser09}. Also less abundant CO isotopologues assume ISM-like abundances, corrected for a fixed [$^{13}$C]/[$^{12}$C] and [$^{18}$O]/[$^{16}$O] ratio. Isotope-selective photodissociation is accounted for empirically by decreasing the C$^{18}$O abundance by an additional factor of 3.
On the other side, as in \cite{Miotello14}, the models presented in this paper use a star-like irradiation source for determining the disk dust and gas temperature. Also, a full chemical calculation is run in order to obtain both a self-consistent gas temperature distribution and atomic and molecular abundances.
\begin{figure*}
   \resizebox{\hsize}{!}
             {\includegraphics[width=1.\textwidth]{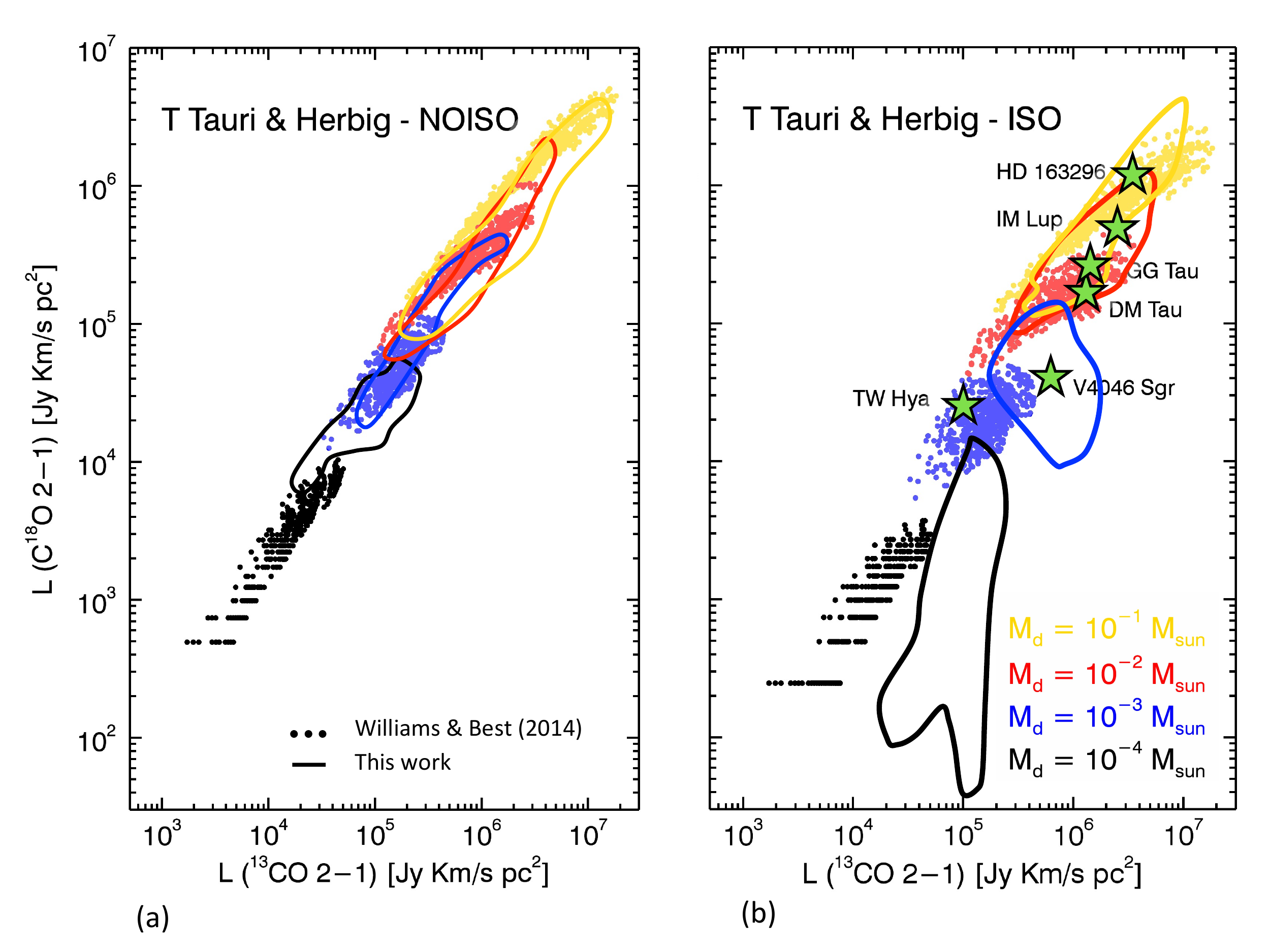}}
      \caption{C$^{18}$O vs $^{13}$CO line luminosities.  Different colors represent models with different disk gas masses. Small dots show the results by \cite{Williams14} and the solid lines outline the regions covered by our model results. Panel (a) shows models where isotope-selective processes are not considered (NOISO). Panel (b) reports models where isotope-selective processes are implemented (ISO) and models by \cite{Williams14}  where the C$^{18}$O abundance has been reduced by a factor of 3. Observations of six well studied disks are reported using the green stars.}
       \label{fig_NOISO}
\end{figure*}

\subsubsection{NOISO models}
\label{NOISO}
First, it is important to compare the two sets of models where isotope-selective processes have not been implemented. This allows us to find differences only due to different methodologies, before analyzing those caused by the effects of isotope-selective photodissociation. For this reason we refer to the line intensities obtained without reducing the C$^{18}$O abundances by a factor of 3 in \cite{Williams14} and compare them with those obtained with the NOISO network. 

Both sets of results are presented in Fig \ref{fig_NOISO} (panel a): small dots show \cite{Williams14} models and solid lines outline the regions covered by the NOISO models. While the \cite{Williams14} models present a clear mass segregation, our model results are more degenerate, in particular for disk masses between $10^{-3}$ and  $10^{-1}$ $M_{\odot}$. This difference is caused by a temperature effect. The higher mass disks presented in our grid are generally colder than those obtained by \cite{Williams14}, even if a 40 $L_{\odot}$ Herbig star is taken. As the disk mass increases, the column densities are enhanced too, deeper regions are colder and a larger fraction of CO is frozen out onto grains. The emitting fraction of CO (isotopologues) is strongly reduced and the line luminosities do not scale linearly with disk mass anymore, in particular for optically thin lines (i.e., C$^{18}$O lines). By contrast, our NOISO line intensities are up to one order of magnitude higher at the low disk mass end.
This is because only few NOISO models with disk masses of $10^{-4}$ $M_{\odot}$ are cold enough to cover the same range of very low line luminosity populated by \cite{Williams14} results. Summarizing, \cite{Williams14} parameterize a wider range of temperatures than in our models and consequently they find a wider range of CO luminosities. For the stellar and disk parameters assumed in this paper, the NOISO models do not reproduce the extreme (both cold and hot) temperature distributions assumed by \cite{Williams14}.

Even though the uncertainties in the gas temperature determination are not important for low-$J$ CO lines (they come from regions in the disk where $T_{\rm gas}$ is coupled to $T_{\rm dust}$), there is a direct connection between the disk structure and the emitted lines. For example, high-$J$ CO lines are detected primarily in flared disks \citep{Meeus13} and can only be reproduced if $T_{\rm gas}$ and $T_{\rm dust}$ are decoupled \citep{Bruderer12}. This aspect is faded out when assuming the same temperature parametrization for all disk structures. Therefore the physical-chemical approach is an improvement compared with the parametric modeling method.

\subsubsection{ISO models}
As shown in panel (b) of Fig. \ref{fig_NOISO}, the implementation of isotope-selective effects in the models affects differently the line luminosities for different disk masses. Less massive disks are more heavily affected than more massive disks, where freeze-out dominates. More precisely, the ISO $^{13}$CO line luminosities are similar to those obtained with the NOISO network, while C$^{18}$O lines are less bright, in particular for low mass disks  ($M_{\rm disk}$ between 10$^{-4}$ and $10^{-3}M_{\odot}$). \cite{Williams14} assume instead a constant reduction of a factor 3 of C$^{18}$O abundances in each mass bin in order to empirically consider isotope-selective effects.
Comparing the two sets of models, a factor of 3 reduction for C$^{18}$O reproduces well the line luminosities for very high mass disks ($M_{\rm d}=10^{-1} M_{\odot}$), but not for lower mass disks, where an additional factor of four would need to be applied. In addition, \cite{Williams14} models do not show such a large degeneracy in the mass determination of very massive disks, as that shown by the ISO models. 

The difference between the two sets of models may be explained also through application to existing data. Observed line luminosities of six well-studied disks reported by \cite{Williams14} are presented in panel (b) of Fig. \ref{fig_NOISO}. In the lower mass regime, where isotope-selective photodissociation is more effective, only two disks are reported (V4046 Sgr and TW Hya) so the comparison is not well representative. For V4046 Sgr there is agreement between the mass derivations obtained by the two grids of models, i.e., $M_{\rm disk}\approx10^{-3}M_{\odot}$. The TW Hya lines are instead just outside the grid of models presented in this paper, while \cite{Williams14} obtain a mass of $\sim 10^{-3}M_{\odot}$. TW Hya is however a peculiar disk, for which a high level of carbon depletion has to be assumed in order to reach agreement with the derived mass from HD observsations \citep[][Kama et al, submitted]{Bergin13,Favre13}. This particular disk will be treated in detail in a future paper. Turning now to the higher mass regime, DM Tau and GG Tau are found to have $\sim 10^{-2}M_{\odot}$ disk masses by \cite{Williams14}, while the ISO models give values that are degenerate by at least one order of magnitude: their mass could be between $10^{-2}M_{\odot}$ and $10^{-1}M_{\odot}$. Finally, the mass of the HD163296 disk is $10^{-1}M_{\odot}$ following \cite{Williams14}, while the ISO models indicate a mass range between $10^{-2}M_{\odot}$ and $10^{-1}M_{\odot}$.

\subsection{Analysis of CO isotopologues observations}
One of the main goals of this paper is to provide outputs that can be employed by the community for analyzing CO isotopologue observations accounting for isotope-selective processes. For this reason integrated line intensities have been computed and tabulated for various isotopologues and different low-$J$ transitions (see Appendix \ref{AppendixA}).

Comparing the observed line intensities with the simulated intensities listed in Table \ref{tab_lineint}, one can find the closest models predictions to the observations and the total disk mass can be constrained, assuming a range of other disk parameters such as $R_{\rm c}, \, \gamma, \, h_{\rm c}, \, \text{and} \, \psi$. Surely, the larger the number of different isotopologues and transitions the more secure the inferred values, although even a single line can be used as a mass tracer in the optically thin regime (Fig.~3).

Knowledge about some disk and stellar parameters helps in identifying the ``best-fit" model. Knowing the stellar luminosity and type narrows down to either the T Tauri or the Herbig disk models. Spatially resolved observations could help in constraining $R_{\rm c}$ and $\gamma$, as explained in Sect, \ref{sp_res} and shown in Fig. \ref{conv}. Finally, any information available about the vertical structure of the disk, for example from infrared data,  assist constraining the range of scale-height values or flaring angles.

Table \ref{tab_lineint} also reports the continuum fluxes at 870 $\mu$m. These are computed assuming an dust opacity $\kappa_{\nu}= 4.2 \rm \, cm^{2}g^{-1}$ at $\lambda=870 \,\mu$m. Dust mass derivations based on these fluxes will therefore scale directly with the chosen dust opacity and on the fixed gas-to-dust ratio, set to 100. 

The significant computation time of the models presented in this paper
does not allow us to highly populate the grid. However, a comparison
of observed integrated line intensities with our models will allow
improved disk mass derivations, accounting for CO isotope-selective
photodissociation and temperature structures.

\subsection{Complementary tracers: [\ion{O}{i}], [\ion{C}{i}], and [\ion{C}{ii}]}
\label{OI}

\begin{figure*}
   \resizebox{\hsize}{!}
             {\includegraphics[width=1.\textwidth]{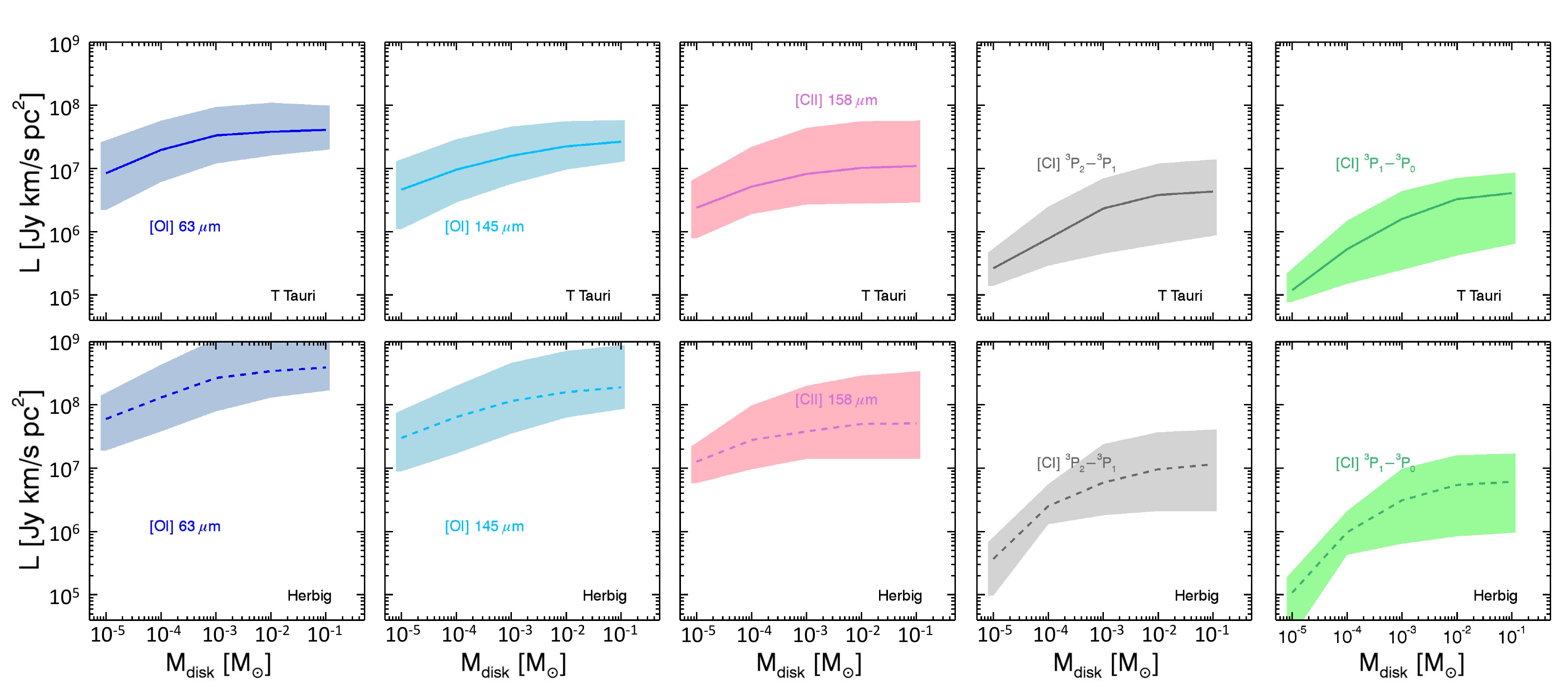}}
      \caption{Median luminosities for [\ion{O}{i}], [\ion{C}{i}] and [\ion{C}{ii}] lines as a function of disk mass. Each line transition is shown by a differently colored line. Solid lines show T Tauri disk models (top panels), while dashed lines present Herbig disk models (bottom panels). The filled regions show the maximum range of the line luminosities obtained with different models.}
       \label{OI_CII_CI}
\end{figure*}

A comparison with complementary disk gas mass tracers is presented in
this section. In particular, we present model results for the atomic
oxygen and atomic and ionized carbon fine-structure transitions. As
shown in Fig.~\ref{OI_CII_CI}, the median luminosities of
[\ion{O}{i}], [\ion{C}{i}], and [\ion{C}{ii}] lines increase only slowly with disk masses. The luminosities vary from a factor of a
few to two order of magnitude over a range of 5 orders of magnitudes
in disk masses. In contrast, Fig.~\ref{L_mass} shows an increase
of CO isotopologues line luminosities of at least four orders of
magnitudes over the same range of disk masses. This comparison
illustrates that CO isotopologues emission is much more sensitive to
total disk mass variations than [\ion{O}{i}], [\ion{C}{i}], and
[\ion{C}{ii}]. For example [\ion{O}{i}] 63 $\mu$m line intensities increase by a factor of a few in our grid, in agreement with what found by \cite{Woitke15} using the PRODIMO code (Fig. 13). Earlier studies using the same code \citep{Kamp10,Kamp11} find that [\ion{O}{i}] 63 $\mu$m line intensities increase by few orders of magnitude over the same range of masses. The reason for the difference between these two results obtained with PRODIMO is not clear.

The [\ion{O}{i}] $^{3}P_1-^{3}P_2$ ($\lambda =$63.2$\mu$m) line has a
high flux in protoplanetary disks. The \emph{Herschel} open time key
program GASPS \citep{Dent13} has detected this line in the majority of
the observed T Tauri disks. Given its high excitation energy (227 K),
[\ion{O}{i}] 63 $\mu$m line arises from the surface layers of the
disk, above the CO low-$J$ emission regions. For this reason,
[\ion{O}{i}] is more sensitive to temperature variations than total
mass variations. The bulk of the mass resides in much deeper regions
of the outer disks, where column densities are higher and CO low-$J$
lines arise from. Accordingly [OI] lines are more sensitive to the disk vertical structure and may help in constraining the scale height and the flaring angle and in reducing the degeneracy in CO-based disk mass determinations. As shown in Fig. \ref{flat_flared}, in our models median [OI] lines are up to one order of magnitude brighter for flared disks ($h_{\rm c}=0.2$, $\psi=0.2$) than for flat disks ($h_{\rm c}=0.1$, $\psi=0.1$). 

While CO is the main gaseous carbon carrier in the warm molecular
layer, neutral and ionized atomic carbon C$^{0}$ and C$^+$
([\ion{C}{i}] and [\ion{C}{ii}] when line emission is concerned)
become the main reservoirs in the UV-irradiated disk surface. The
optically thin lines of these species are a direct probe of the carbon
abundance in the disk surface \citep[][Kama et al. submitted]{Kama16}. However,
[\ion{C}{ii}] cannot be observed from the ground and the data from
\emph{Herschel} may be contaminated by extended emission
\citep{Fedele13,Dent13}. [\ion{C}{i}] $^{3}P_1-^{3}P_0$ emission at 492 GHz and  [\ion{C}{i}] $^{3}P_2-^{3}P_1$ emission at 809 GHz has recently been
unambiguously detected with the \emph{Atacama Pathfinder EXperiment}
(APEX) telescope in several disks, and used to constrain the gas-phase
carbon abundance, since it is not sensitive to the temperature structure \citep[][Kama et al. submitted]{Kama16}.

\subsection{Effects of lower carbon abundance}
\label{C_depletion}

The abundance of CO and its isotopologues depends on the total
abundance of elemental gas-phase carbon, which may be lower than in
the ISM ([C]/[H]$_{\rm ISM} = 1.35 \times 10^{-4}$) as suggested by
recent observations and analyses
\citep[e.g.,][Kama et al. submitted]{Bruderer12,Favre13,Kama16}. Thus, there is a
degeneracy between the carbon abundance and the total gas mass
inferred from CO isotopologues.  To quantify this effect, we ran ten T Tauri disk models with a reduced initial carbon abundance. The standard value
assumed for the whole grid, $X_{\rm C}=\text{[C]/[H]}_{\rm ISM}=1.35
\times 10^{-4}$, has been reduced by factors of 3 and 10. The total
abundance of $^{13}$C has been reduced by the same factors. The other
disk model parameters are kept at $R_{\rm c}=60$ au, $\gamma$=0.8, $h_{\rm c}$=0.1, $\psi$=0.1, and $M_{\rm disk}=10^{-5},10^{-4},10^{-3},10^{-2},10^{-1} M_{\odot}$. Both the oxygen and the PAHs abundances are kept as in the large grid models (see Table \ref{Tab_abundances} and Table \ref{tab:modelpar}).

The results are shown in Fig.~\ref{C_depl} and quantified in Appendix
\ref{AppendixB}.  As expected, both the $^{13}$CO and C$^{18}$O line
intensities are reduced if the initial carbon abundance is
lowered. For the more massive disks, the reduction is larger for the
more optically thin C$^{18}$O than for $^{13}$CO lines. This difference
decreases for the lower-mass disks, where both lines scale more
linearly with [C]/[H]$_{\rm gas}$. This trend is similar to that
resulting from a reduction in disk mass, as shown in
Fig.~\ref{clouds}. This is to be expected from the direct dependence
of CO isotopolog abundances on both [C]/[H]$_{\rm gas}$ and the total
gas mass.

\begin{figure*}
   \resizebox{\hsize}{!}
             {\includegraphics[width=1.\textwidth]{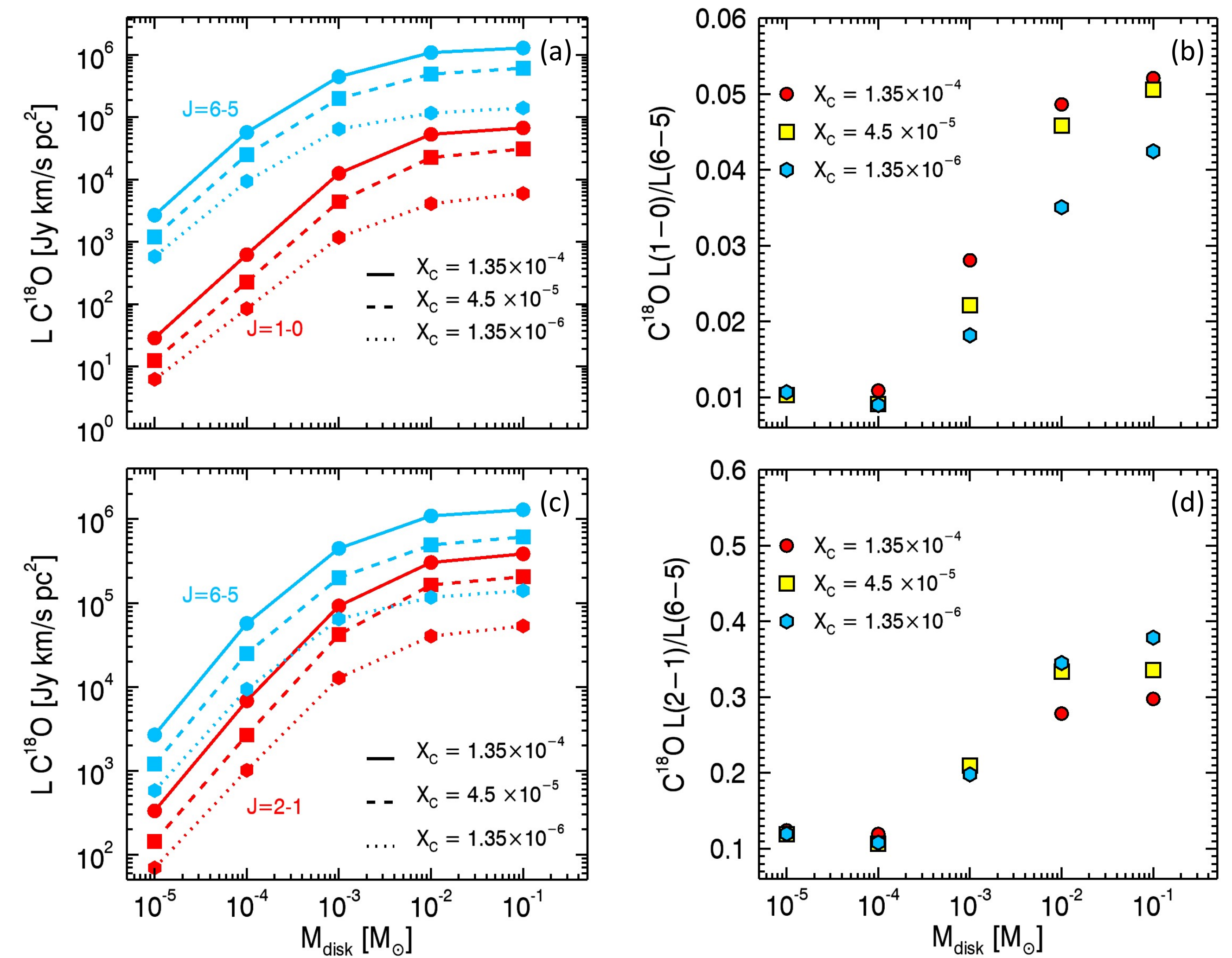}}
      \caption{Results of T Tauri disk models with carbon abundances reduced by factors of 3 and 10. Panel (a) shows C$^{18}$O $J=1-0$ and $J=6-5$ line luminosities as a function of the disk mass (respectively in red and blue), and panel (c)
the C$^{18}$O $J=2-1$ and $J=6-5$ line luminosities. The models with ISM-like carbon abundances are shown by the solid lines, those with a factor of 3 of carbon depletion by the dashed lines and those with a factor of 10 depletion by the dotted lines. Panel (b) shows C$^{18}$O $J=1-0/J=6-5$ line luminosity ratios as a function of the disk mass, and panel (d) the $J=2-1/J=6-5$ ratios.  The models with ISM-like carbon abundances are shown by the red circles, those with a factor of 3 of carbon depletion by yellow squares and those with a factor of 10 depletion by the blue hexagons. Model disk parameters: $R_{\rm c}=60$ au, $\gamma$=0.8, $h_{\rm c}$=0.1, $\psi$=0.1, and $M_{\rm disk}=10^{-5},10^{-4},10^{-3},10^{-2},10^{-1} M_{\odot}$.}
       \label{C_depl_ratios}
\end{figure*}

An underabundace of gas-phase carbon could show up as extended scatter
in Fig.~\ref{clouds}, especially along the C$^{18}$O axis which owing
to its optically thin lines is expected to be more directly sensitive
to the number of available carbon atoms.  The behaviour of C$^{18}$O
emission with carbon abundance is shown for a subset of models in
Fig. \ref{C_depl_ratios}, panels (a) and (c).  We also investigated
whether ratios of different-$J$ transitions of a single isotopologue,
C$^{18}$O, can brake the abundance-mass degeneracy.  Panel (b) and (d)
of Fig. \ref{C_depl_ratios} present the model C$^{18}$O $J=1-0/6-5$
and $J=2-1/6-5$ line luminosity ratios as a function of the disk mass
for different carbon abundances.  These four figures demonstrate that the
relations are more complex than simply linear: in some regimes 
C$^{18}$O scales sub-linearly, in others super-linearly with carbon
abundance. The explanation for these trends is further discussed in Appendix
\ref{AppendixB}.

Concluding, a combination of different CO isotopologues (e.g.,
$^{13}$CO and C$^{18}$O) and various $J$ transitions (e.g., C$^{18}$O
$J=2-1$ and $J=6-5$) together with [\ion{C}{i}] observations would
provide stronger observational constraints on the total disk gas mass,
because this could also provide some indication on the level of carbon
depletion.  This result has to be seen as indicative, however, because
the models with a lower carbon abundances presented here only cover a small range of disk
parameters. A more extended grid of models is needed to quantify the
predictions on line ratios between various $J$-transitions of the same
isotopolog.

\section{Summary and Conclusion}

\label{Summary}
We presented a grid of 840 disk models where CO isotope-selective photodissociation has been treated in a complete way. 
A full thermo-chemical model was run and abundances and line intensities were obtained as outputs. The main conclusions are listed below.
\begin{itemize}
\item When CO isotope selective photodissociation is considered, the abundances of CO isotopologues are affected. In particular, there are regions in the disk where $\rm C^{18}O$ and  $\rm C^{17}O$ show an underabundance of more than one order of magnitude with respect to  $\rm ^{12}CO$ when compared with the overall elemental abundance ratios. $\rm ^{13}CO$ line intensities are instead slightly enhanced \citep{Miotello14}.

\item Observations of more than one CO-isotopologue, such as $\rm ^{13}CO$ and $\rm C^{18}O$, can be employed for determining disk total masses. There is however some degeneracy,  of order one order of magnitude, for disks more massive than $10^{-3} M_{\odot}$. In the low-mass optically thin regime, where the line intensities scale linearly with disk mass, $\rm ^{13}CO$ or $\rm C^{18}O$ alone can be employed as mass tracers and simple fit relations are presented. These species remain robust gas-mass tracers even for much lower gas-to-dust ratios.

\item Constraining $R_{\rm c}$ and $\gamma$ with spatially resolved line observations reduces the degeneracy in disk mass determinations by a factor between 2 and 5.

\item Comparison of results of our grid with parametric models shows that our $\rm ^{13}CO$ and $\rm C^{18}O$ line intensities computed with the NOISO network are up to one order of magnitude higher at the low disk mass end than those obtained by \cite{Williams14}. Moreover, in our models isotope-selective effect are dependent on disk mass: the factor of 3 reduction for $\rm C^{18}O$ assumed by \cite{Williams14} does not reproduce the line intensities for masses lower than $10^{-3} M_{\odot}$, where an additional factor of four would need to be applied.

\item CO isotopologues emission is more sensitive to total disk mass variations than that of [\ion{O}{i}], [\ion{C}{i}], and
[\ion{C}{ii}]. These lines can in fact be employed to constrain the disk vertical structure.

\item The degeneracy with carbon abundance has been investigated for a small set of models. [\ion{C}{i}] observations give some indication on the level of carbon depletion and provide stronger constraints on the disk mass determination if combined with CO isotopologues observations.  

\item Finally, different CO isotopologues total fluxes for various low-$J$ transitions are provided in Appendix A. They can be used for the analysis of CO isotopologue observations, taking into account the isotope-selective effects.
\end{itemize}

\section*{Acknowledgements}

The authors are grateful to Jonathan Williams, Catherine Walsh, Paolo Cazzoletti, Leonardo Testi and the anonymous referee for useful discussions and comments.
Astrochemistry in Leiden is supported by the Netherlands Research
School for Astronomy (NOVA), by a Royal Netherlands Academy of Arts
and Sciences (KNAW) professor prize, and by the European Union A-ERC
grant 291141 CHEMPLAN.



\clearpage
\begin{appendix}

\section{Additional tables and figures}
\label{AppendixA}

\begin{table}[h!]
\caption{Polynomial coefficients $A_{y}$, $B_{y}$, $C_{y}$, and $D_{y}$, in Eq. (\ref{L_13})}
\label{A_mass}
\centering
\begin{tabular}{lcc}
\toprule
&T Tauri&Herbig\\
\cmidrule(lr){2-3}
$A_{13}$&$-8.954 \cdot 10^{2}$&$-8.746 \cdot 10^{4}$\\
$B_{13}$&$1.477 \cdot 10^{9}$&$1.245 \cdot 10^{9}$\\
$C_{13}$&$2.954 \cdot 10^{6}$&$5.214\cdot 10^{6}$\\
$D_{13}$&$7.201 \cdot 10^{5}$&$1.402\cdot 10^{6}$\\
$M_{\rm tr} [M_{\odot}]$&$2 \cdot 10^{-4}$&$5\cdot 10^{-4}$\\
\cmidrule(lr){2-3}
$A_{18}$&$-2.852 \cdot 10^{2}$&$-8.026 \cdot 10^{2}$\\
$B_{18}$&$5.875 \cdot 10^{7}$&$8.571 \cdot 10^{7}$\\
$C_{18}$&$1.235 \cdot 10^{6}$&$2.276\cdot 10^{6}$\\
$D_{18}$&$4.180 \cdot 10^{5}$&$8.001\cdot 10^{5}$\\
$M_{\rm tr} [M_{\odot}]$&$2.5 \cdot 10^{-3}$&$2.5 \cdot 10^{-3}$\\
\hline
\end{tabular}
\end{table}

\begin{table}[h!]
\caption{Polynomial coefficients $A_{R_{\rm c}}$, $B_{R_{\rm c}}$, $C_{R_{\rm c}}$, and $D_{R_{\rm c}}$ in Eq. (\ref{polin}). }
\label{poly_coeff}
\centering
\begin{tabular}{lcc}
\toprule
&T Tauri&Herbig\\
\cmidrule(lr){2-3}
$A_{30}$&$1.329 \cdot 10^{-2}$&$1.008\cdot 10^{-2}$\\
$B_{30}$&$7.503 \cdot 10^{2}$&$4.730 \cdot 10^{2}$\\
$C_{30}$&$4.901 \cdot 10^{-1}$&$5.305\cdot 10^{-1}$\\
$D_{30}$&$8.724 \cdot 10^{-2}$&$1.000\cdot 10^{-1}$\\
$M_{\rm tr} [M_{\odot}]$&$2 \cdot 10^{-4}$&$4 \cdot10^{-4}$\\
\cmidrule(lr){2-3}
$A_{60}$&$1.511 \cdot 10^{-2}$&$8.492 \cdot 10^{-3}$\\
$B_{60}$&$2.234 \cdot 10^{2}$&$1.448 \cdot 10^{2}$\\
$C_{60}$&$4.873 \cdot 10^{-1}$&$5.934\cdot 10^{-1}$\\
$D_{60}$&$1.117 \cdot 10^{-1}$&$1.550\cdot 10^{-1}$\\
$M_{\rm tr} [M_{\odot}]$&$5 \cdot 10^{-4}$&$5 \cdot10^{-4}$\\
\cmidrule(lr){2-3}
$A_{200}$&$1.141 \cdot 10^{-2}$&$1.026 \cdot 10^{-2}$\\
$B_{200}$&$5.104 \cdot 10^{1}$&$1.662 \cdot 10^{1}$\\
$C_{200}$&$5.561 \cdot 10^{-1}$&$7.109\cdot 10^{-1}$\\
$D_{200}$&$1.640 \cdot 10^{-1}$&$2.250\cdot 10^{-1}$\\
$M_{\rm tr} [M_{\odot}]$&$2 \cdot 10^{-3}$&$2 \cdot10^{-3}$\\
\hline
\end{tabular}
\end{table}

\begin{table}[h!]
\label{chemspec}
\caption{Species contained in the ISO chemical network. \emph{Notation}: H$_2^*$ refers to vibrationally excited H$_2$;  PAH$^0$, PAH$^+$ and PAH$^-$ are neutral, positively and negatively charged PAHs, while PAH:H denotes hydrogenate PAHs; JX refers to species frozen-out onto dust grainis.}
\centering
{\footnotesize
\begin{tabular}{ccccc}
\hline\hline
H        &  He       &  C        &  $^{13}$C       &  N  \\       
O        &  $^{17}$O      &  $^{18}$O      &  Mg        &  Si      \\  
S        &  Fe       &  H$_{2}$       &  H$_{2}$$^{*}$       &  CH     \\   
$^{13}$CH     &  CH$_{2}$      &  $^{13}$CH$_{2}$    &  NH        &  CH$_{3}$  \\     
$^{13}$CH$_{3}$    &  NH$_{2}$      &  CH$_{4}$      &  $^{13}$CH$_{4}$     &  OH        \\
$^{17}$OH     &  $^{18}$OH     &  NH$_{3}$      &  H$_{2}$O       &  H$_{2}$$^{17}$O     \\
H$_{2}$$^{18}$O    &  CO       &  C$^{17}$O     &  C$^{18}$O      &  $^{13}$CO      \\
$^{13}$C$^{17}$O   &  $^{13}$C$^{18}$O   &  HCN      &  H$^{13}$CN     &  HCO       \\
HC$^{17}$O    &  HC$^{18}$O    &  H$^{13}$CO    &  H$^{13}$C$^{17}$O   &  H$^{13}$C$^{18}$O   \\
NO       &  N$^{17}$O     &  N$^{18}$O     &  H$_{2}$CO      &  H$_{2}$C$^{17}$O    \\
H$_{2}$C$^{18}$O   &  H$_{2}$$^{13}$CO   &  H$_{2}$$^{13}$C$^{17}$O &  H$_{2}$$^{13}$C$^{18}$O  &  O$_{2}$     \\   
O$^{17}$O     &  O$^{18}$O     &  CO$_{2}$      &  CO$^{17}$O     &  CO$^{18}$O     \\
$^{13}$CO$_{2}$    &  $^{13}$CO$^{17}$O  &  $^{13}$CO$^{18}$O  &  CN        &  $^{13}$CN      \\
N$_{2}$       &  SiH      &  SiO      &  Si$^{17}$O     &  Si$^{18}$O     \\
H$^{+}$       &  H$^{-}$       &  H$_{2}$$^{+}$      &  H$_3^{+}$       &  He$^{+}$       \\
HCO$^{+}$     &  HC$^{17}$O$^{+}$   &  HC$^{18}$O$^{+}$   &  H$^{13}$CO$^{+}$    &  H$^{13}$C$^{17}$O$^{+}$  \\
H$^{13}$C$^{18}$O$^{+}$ &  C$^{+}$       &  $^{13}$C$^{+}$     &  CH$^{+}$       &  $^{13}$CH$^{+}$     \\
N$^{+}$       &  CH$_{2}$$^{+}$     &  $^{13}$CH$_{2}$$^{+}$   &  NH$^{+}$       &  CH$_3^{+}$      \\
$^{13}$CH$_3^{+}$   &  O$^{+}$       &  $^{17}$O$^{+}$     &  $^{18}$O$^{+}$      &  NH$_{2}$$^{+}$      \\
CH$_{4}$$^{+}$     &  $^{13}$CH$_{4}$$^{+}$   &  OH$^{+}$      &  $^{17}$OH$^{+}$     &  $^{18}$OH$^{+}$     \\
NH$_{3}$$^{+}$     &  CH$_5^{+}$     &  $^{13}$CH$_5^{+}$   &  H$_{2}$O$^{+}$      &  H$_{2}$$^{17}$O$^{+}$    \\
H$_{2}$$^{18}$O$^{+}$   &  H$_3$O$^{+}$     &  H$_3^{17}$O$^{+}$   &  H$_3^{18}$O$^{+}$    &  Mg$^{+}$       \\
CN$^{+}$      &  $^{13}$CN$^{+}$    &  HCN$^{+}$     &  H$^{13}$CN$^{+}$    &  Si$^{+}$       \\
CO$^{+}$      &  C$^{17}$O$^{+}$    &  C$^{18}$O$^{+}$    &  $^{13}$CO$^{+}$     &  $^{13}$C$^{18}$O$^{+}$   \\
HCNH$^{+}$    &  H$^{13}$CNH$^{+}$  &  SiH$^{+}$     &  NO$^{+}$       &  N$^{17}$O$^{+}$     \\
N$^{18}$O$^{+}$    &  SiH$_{2}$$^{+}$    &  S$^{+}$       &  O$_{2}$$^{+}$       &  O$^{17}$O$^{+}$     \\
O$^{18}$O$^{+}$    &  SiO$^{+}$     &  Si$^{17}$O$^{+}$   &  Si$^{18}$O$^{+}$    &  CO$_{2}$$^{+}$      \\
CO$^{17}$O$^{+}$   &  CO$^{18}$O$^{+}$   &  $^{13}$CO$_{2}$$^{+}$   &  $^{13}$CO$^{17}$O$^{+}$  &  $^{13}$CO$^{18}$O$^{+}$  \\
Fe$^{+}$      &  SiOH$^{+}$    &  Si$^{17}$OH$^{+}$  &  Si$^{18}$OH$^{+}$   &  HN$_{2}$$^{+}$      \\
e$^{-}$       &  PAH$^0$     &  PAH$^{+}$     &  PAH$^{-}$      &  PAH:H     \\
JC       &  J$^{13}$C     &  JN       &  JO        &  J$^{17}$O      \\
J$^{18}$O     &  JCH      &  J$^{13}$CH    &  JCH$_{2}$      &  J$^{13}$CH$_{2}$    \\
JNH      &  JCH$_{3}$     &  J$^{13}$CH$_{3}$   &  JNH$_{2}$      &  JCH$_{4}$      \\
J$^{13}$CH$_{4}$   &  JOH      &  J$^{17}$OH    &  J$^{18}$OH     &  JNH$_{3}$     \\
JH$_{2}$O     &  JH$_{2}$$^{17}$O   &  JH$_{2}$$^{18}$O   &  JCO       &  JC$^{17}$O   \\  
JC$^{18}$O    &  J$^{13}$CO    &  J$^{13}$C$^{17}$O  &  J$^{13}$C$^{18}$O   &  JCO$_{2}$      \\
JCO$^{17}$O   &  JCO$^{18}$O   &  J$^{13}$CO$_{2}$   &  J$^{13}$CO$^{17}$O  &  J$^{13}$CO$^{18}$O  \\
\hline
\end{tabular}
}
\end{table} 

Tables \ref{A_mass} and \ref{poly_coeff} report the polynomial coefficients $A^{13}$ and $A^{18}$ for $L_{13}$ and $L_{18}$ in Eq. (\ref{L_13}) and the polynomial coefficients $A_{R_{\rm c}}$ for $L_{18}/L_{13}$ in Eq. (\ref{polin}).

Table \ref{tab_lineint} reports $^{12}$CO, $^{13}$CO, C$^{18}$O, C$^{17}$O and [\ion{C}{i}] total fluxes (in K km/s) simulated by the models presented in this paper. The fluxes are computed assuming a distance of 100 pc and a beam size of 10". CO isotopologues fluxes are computed for four molecular line transitions ($J=1-0, J=2-1, J=3-2, J=6-5$) and for two disk inclination angles ($i=10^{\circ},80^{\circ}$). This table, together with the analogue one for Herbig disk models, is provided on-line in ascii format.

Table. A.4 reports the list of all the species considered in the ISO chemical network.

\begin{figure*}
   \resizebox{\hsize}{!}
             {\includegraphics[width=1.\textwidth]{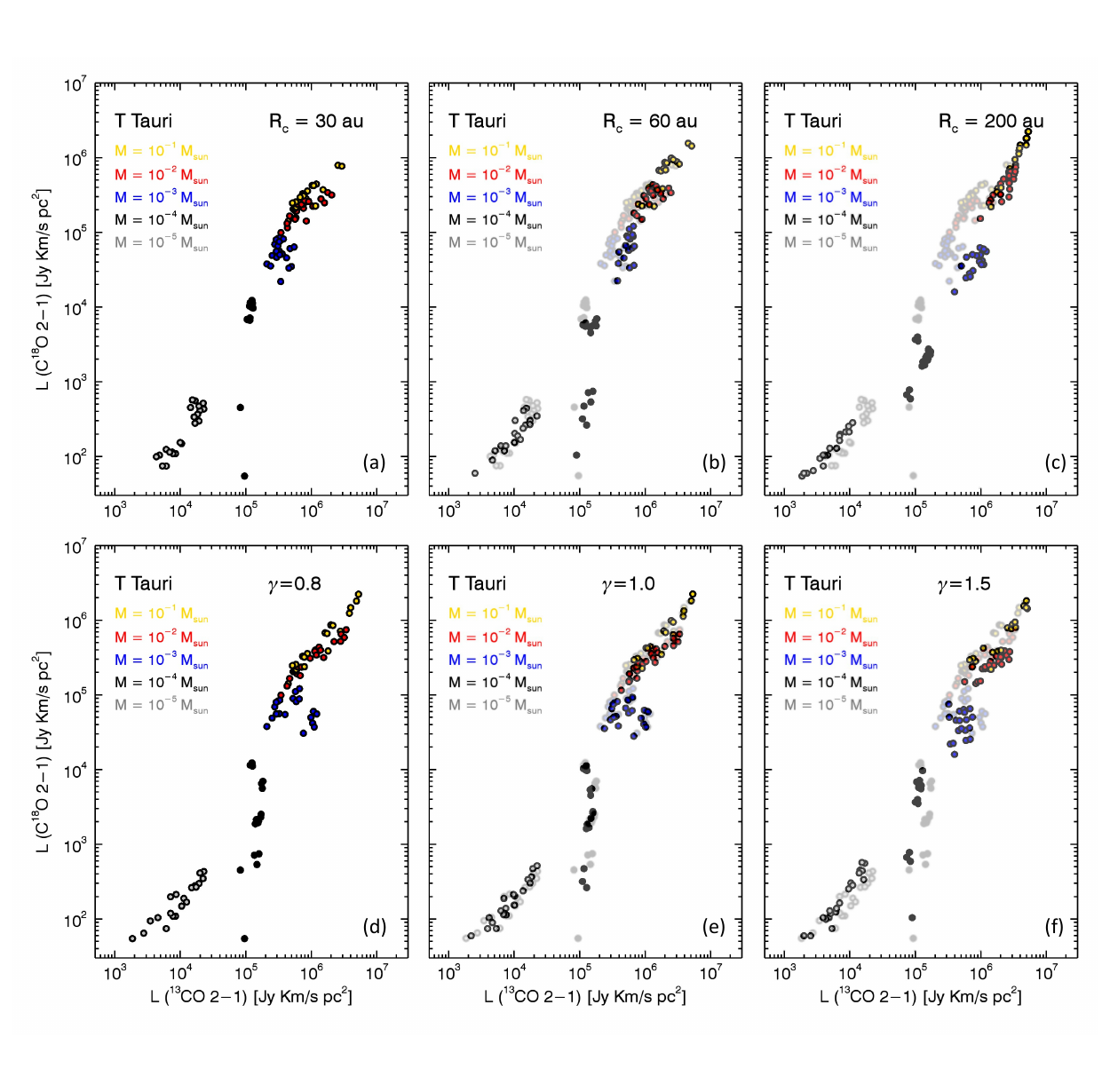}}
      \caption{C$^{18}$O vs $^{13}$CO ($J=2-1$) line luminosity obtained implementing isotope-selective processes. Different colors represent models with different disk gas masses. In the upper panels models with different values for $R_{\rm c}$ are presented separately. In the lower panels models with different values for $\gamma$ are shown separately. Models displayed in panel (a) and (d) are also reported in gray in panels (b,c) and (e,f) respectively in order to facilitate the comparison.}
       \label{param}
\end{figure*}

\begin{figure}[h!]
   \resizebox{\hsize}{!}
             {\includegraphics[width=1.\textwidth]{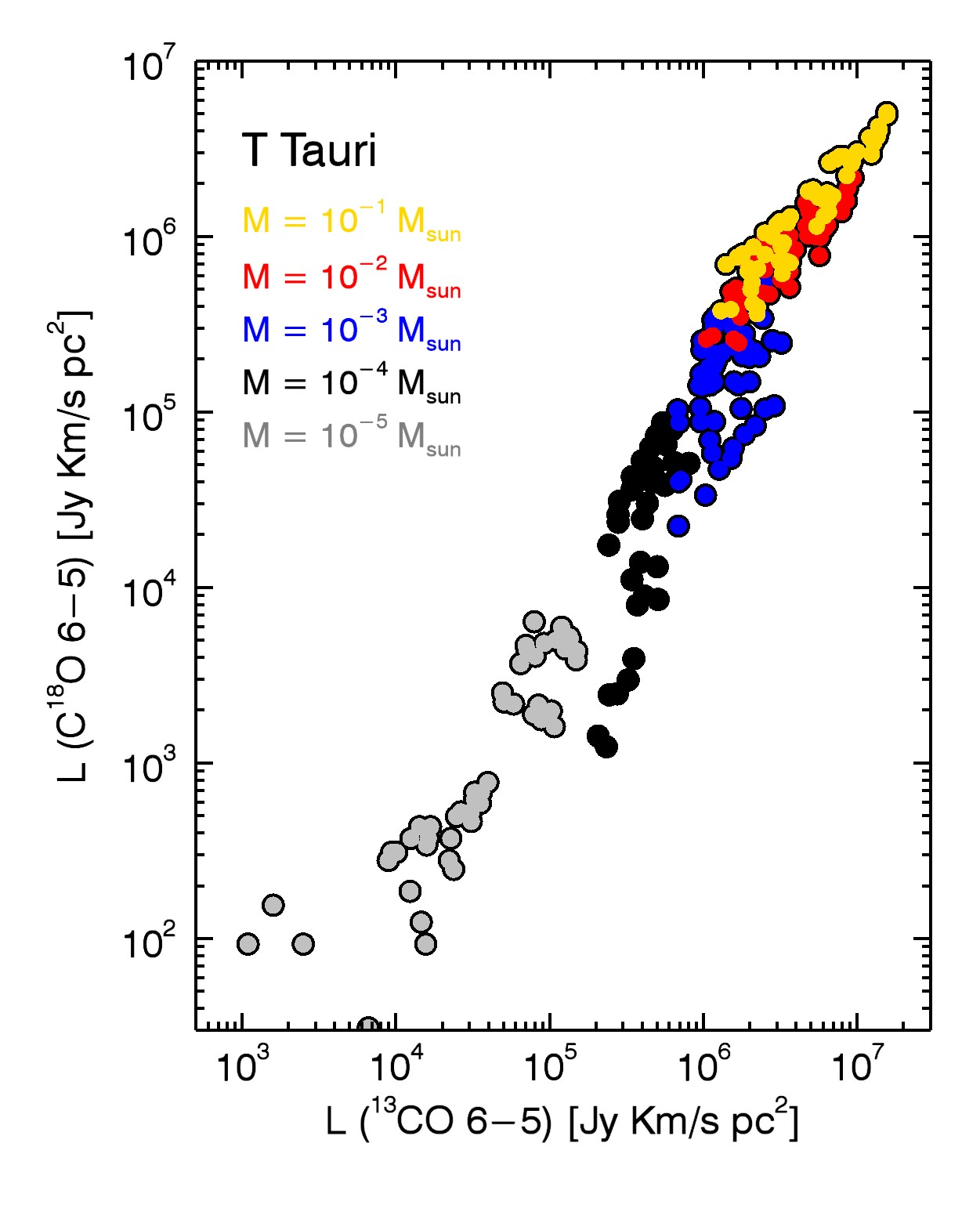}}
      \caption{C$^{18}$O vs $^{13}$CO ($J=6-5$) line luminosity obtained implementing isotope-selective processes. Different colors represent models with different disk gas masses.}
       \label{56}
\end{figure}

\begin{figure*}[tbh]
\centering
   \resizebox{.7\textwidth}{!}
            {\includegraphics[width=0.5\textwidth]{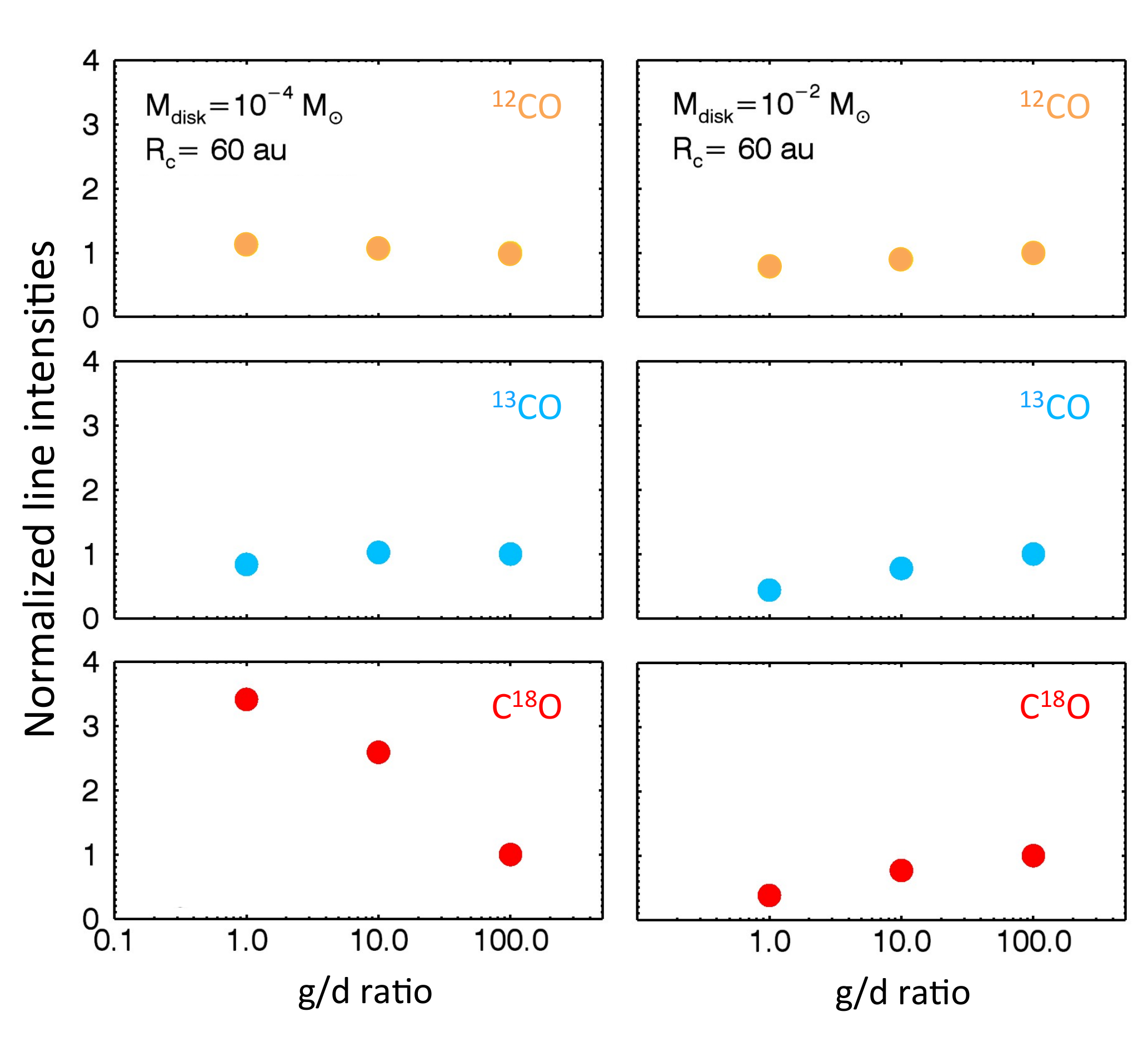}}
      \caption{CO isotopologues line intensities normalized by the intensity for gas-to-dust=100 as a function of gas-to-dust ratio. Keeping the gas mass fixed, two T Tauri disk models ($R_{\rm c}=60$ au, $\gamma=1$, $\psi=0.2$, $h_{\rm c}=0.2$, $M_{\rm disk}=10^{-4}, 10^{-2}M_{\odot}$) have been run with gas-to-dust mass ratios of 100, 10 and 1. Orange, blue and red circles show respectively $^{12}$CO, $^{13}$CO, and C$^{18}$O lines.} 
       \label{check_g_d}
\end{figure*}

\begin{figure}[h!]
   \resizebox{\hsize}{!}
            {\includegraphics[width=1.\textwidth]{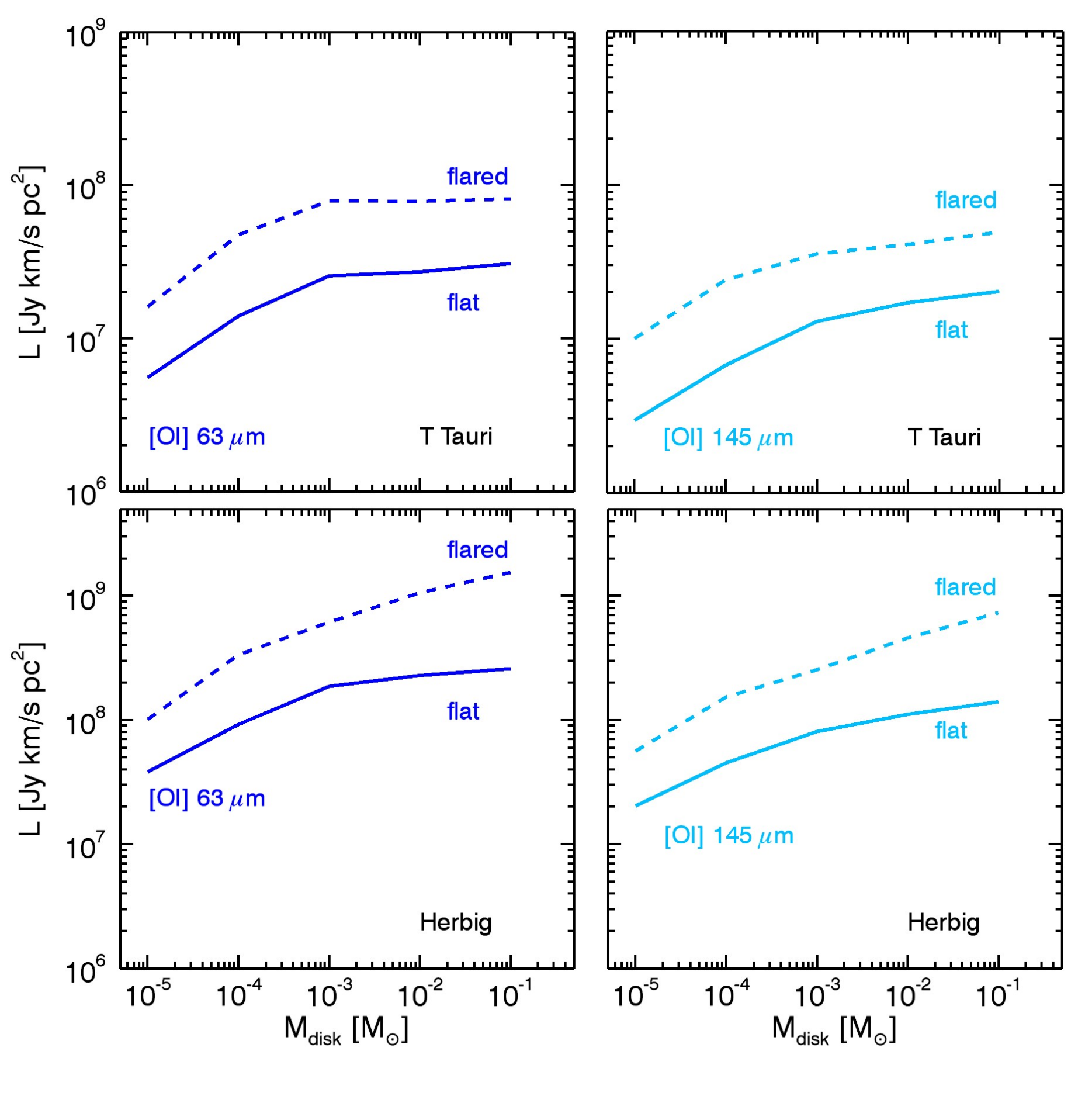}}
      \caption{Median luminosities for [\ion{O}{i}] lines as a function of disk mass for flat ($h_{\rm c}=0.1$, $\psi=0.1$) and flared ($h_{\rm c}=0.2$, $\psi=0.2$) disks, shown by the solid and the dashed lines respectively. T Tauri disk models are shown in upper panels and Herbig models are presented in the lower panels.}
       \label{flat_flared}
\end{figure}

Finally, additional figures are also reported in this appendix. Fig. \ref{param} shows C$^{18}$O vs $^{13}$CO ($J=2-1$) line luminosity obtained varying different disk parameters, such as $R_{\rm c}$ and $\gamma$. Fig. \ref{56} shows C$^{18}$O vs $^{13}$CO ($J=6-5$) line luminosity obtained implementing isotope-selective processes. Fig. \ref{flat_flared} shows the median luminosities for [\ion{O}{i}] lines as a function of disk mass for flat ($h_{\rm c}=0.1$, $\psi=0.1$) and flared ($h_{\rm c}=0.2$, $\psi=0.2$) disk models. Fig. \ref{check_g_d} presents CO isotopologues line intensities as a function of gas-to-dust ratio. The same T Tauri disk model ($R_{\rm c}=60$ au, $\gamma=1$, $\psi=0.2$, $h_{\rm c}=0.2$, $M_{\rm disk}=10^{-4}M_{\odot}$) has been run with gas-to-dust mass ratios of 100, 10 and 1.

\begin{table*}[tbh]
\caption{The first five columns show the parameters assumed for each T Tauri disk model in our large grid. Then, the continuum emission at 870 $\mu$m in K, $^{12}$CO, $^{13}$CO, C$^{18}$O, C$^{17}$O and [CI] total fluxes simulated by these models are reported, expressed in K km/s. The fluxes are computed assuming a distance of 100 pc and a beam size of 10". CO isotopologues fluxes are computed for four molecular line transitions ($J=1-0, J=2-1, J=3-2, J=6-5$) and for two disk inclination angles ($i=10^{\circ},80^{\circ}$). This table, together with the analogue one for Herbig disk models, is provided on-line in electronic format.}
\label{tab_lineint}
\centering
\begin{tabular}{ccccccccccccccc}
\toprule
$R_{\rm c}$&$\gamma$&$h_{\rm c}$&$\psi$&$M_{\rm disk}$&
$I[\rm 870 \mu m]_{10}$&
$I[^{12}\rm CO]^{1-0}_{10}$&...&
$I[^{13}\rm CO]^{2-1}_{80}$&...&
$I[\rm C^{18}O]^{3-2}_{10}$&...&
$I[\rm C^{17}O]^{6-5}_{80}$&...&$
I[\rm CI]^{2-1}_{80}$\\
\midrule
30.0&0.8&0.1&0.1&$10^{-5}$
&$1.50^{(-4)}$
&$3.16^{(-1)}$&
&$2.21^{(-2)}$&
&$1.46^{(-3)}$&
&$3.00^{(-5)}$&
&$1.94^{(-2)}$\\
30.0&0.8&0.1&0.1&$10^{-4}$
&$1.01^{(-3)}$
&$7.27^{(-1)}$&
&$1.66^{(-1)}$&
&$2.95^{(-2)}$&
&...&&\\
...&&&&&&&&&&&&&\\
\hline
\end{tabular}
\end{table*}

\section{Effects of carbon depletion on CO isotopologue line intensities}
\label{AppendixB}
The elemental composition assumed at the start of the calculation for the 800 models presented in this work is the same as in \cite{Miotello14} and is reported in Table \ref{Tab_abundances}.
In this section, the impact of carbon depletion on CO isotopologues
line luminosities is further quantified.  Ten disk models have been
run with carbon abundances reduced by factors of 3 and 10
compared with the total gas-phase elemental carbon abundance
([C]/[H]$_{\rm gas}=1.35\times 10 {-4}$ (both the ISO and the NOISO networks). 
By averaging the change of fluxes seen in these models, we define a
reduction factor, $\Lambda$, for the CO isotopolog line
intensities. More specifically, if the carbon abundance is reduced by
a factor of 3 or 10 the CO isotopologues line luminosity are defined
respectively as $L(^{13}$CO)$_3$ and $L$(C$^{18}$O)$_3$,
$L(^{13}$CO)$_{10}$ and $L$(C$^{18}$O)$_{10}$. The line luminosities
obtained with a fiducial ISM-like carbon abundance are instead defined
as: $L(^{13}$CO) and $L$(C$^{18}$O). They are related to each
other as follows:
\begin{equation}
\begin{split}
L(^{13}\text{CO})& = \Lambda^{13}_{3}\cdot L(^{13}\text{CO})_3, \\
L(\rm C^{18}O)& = \Lambda^{18}_{3}\cdot L(\rm C^{18}O)_3,\\
L(^{13}\text{CO})& = \Lambda^{13}_{10}\cdot L(^{13}\text{CO})_{10},\\
L(\rm C^{18}O)& = \Lambda^{18}_{10}\cdot L(\rm C^{18}O)_{10},
\end{split}
\end{equation}
where the reduction factors $\Lambda$ are reported in Table \ref{C_red_fact}.  In practice, if the carbon abundance is reduced by factors 3 or 10, CO isotopologues line luminosities are lowered by factors of $ \Lambda_{3}$ or $ \Lambda_{10}$. The superscripts 13 and 18 stand for $^{13}$CO and C$^{18}$O line luminosities.

\begin{table}[tbh]
\caption{Elelmental composition assumed for the calculation. \emph{a(b)} means $a \times 10^b$.}
\label{Tab_abundances}
\centering
\begin{tabular}{lc}
\hline\hline
Element & Number fraction \\
\hline
H& 1  \\   
He& 7.59(-2)\\     
C & 1.35(-4)$^*$\\
$^{13}$C& 1.75(-3) \\   
N&  2.14(-5)\\    
O&  2.88(-4)\\    
$^{17}$O& 1.61(-7)\\    
$^{18}$O& 5.14(-7)\\    
Mg& 4.17(-7)\\   
Si& 7.94(-6)\\    
S& 1.91(-6)\\   
Fe&  4.27(-7)\\
\hline
\end{tabular}
\newline
$^*$ \small{C abundance has been reduced by factors of\\ 3 and 10 for ten disk models (See App. B).}
\end{table}

\begin{table}[tbh]
\caption{Carbon depletion reduction factors that relates CO isotopologues line luminosities (for three transitions: $J=2-1, 1-0, 6-5$) obtained with ISM-like or reduced carbon abundances. In practice, if the carbon abundance is reduced by factors 3 or 10, CO isotopologues line luminosities are lowered by factors of $ \Lambda_{3}$ or $ \Lambda_{10}$. The superscripts 13 and 18 stand for $^{13}$CO and C$^{18}$O.}
\label{C_red_fact}
\centering
\begin{tabular}{rccccc}
\toprule
$M_{\rm disk}\, [M_{\odot}]$&$10^{-5}$&$10^{-4}$&$10^{-3}$&$10^{-2}$&$10^{-1}$\\
\cmidrule(lr){2-6}
&\multicolumn{5}{c}{$J=2-1$}\\
$\Lambda^{13}_{3}$&1.5&1.2&1.1&1.2 &1.1\\
$\Lambda^{18}_{3}$&1.0 &0.9&1.8&1.3&1.7\\
$\Lambda^{13}_{10}$&4.9&3.4&3.1 &3.9&2.3\\
$\Lambda^{18}_{10}$&2.0&2.5 &5.2&5.3 & 5.2\\
\cmidrule(lr){2-6}
&\multicolumn{5}{c}{$J=3-2$}\\
$\Lambda^{13}_{3}$&3.2&2.3& 1.9&2.2&1.9\\
$\Lambda^{18}_{3}$&2.3 &1.7&2.9& 2.3 &2.7\\
$\Lambda^{13}_{10}$&10.3& 6.5& 4.5&6.8&3.6\\
$\Lambda^{18}_{10}$&4.5&3.7 &7.9&4.5&7.8\\
\cmidrule(lr){2-6}
&\multicolumn{5}{c}{$J=6-5$}\\
$\Lambda^{13}_{3}$&13.8&7.1 &4.7&5.1&4.7\\
$\Lambda^{18}_{3}$&6.1 &3.4&4.1&5.3&4.7\\
$\Lambda^{13}_{10}$&40.1& 20.6 &10.4 &14.2 &10.4\\
$\Lambda^{18}_{10}$&10.2 &10.2&13.4&18.4&16.7\\
\hline
\end{tabular}
\end{table}

\begin{figure*}
   \resizebox{\hsize}{!}
             {\includegraphics[width=1.\textwidth]{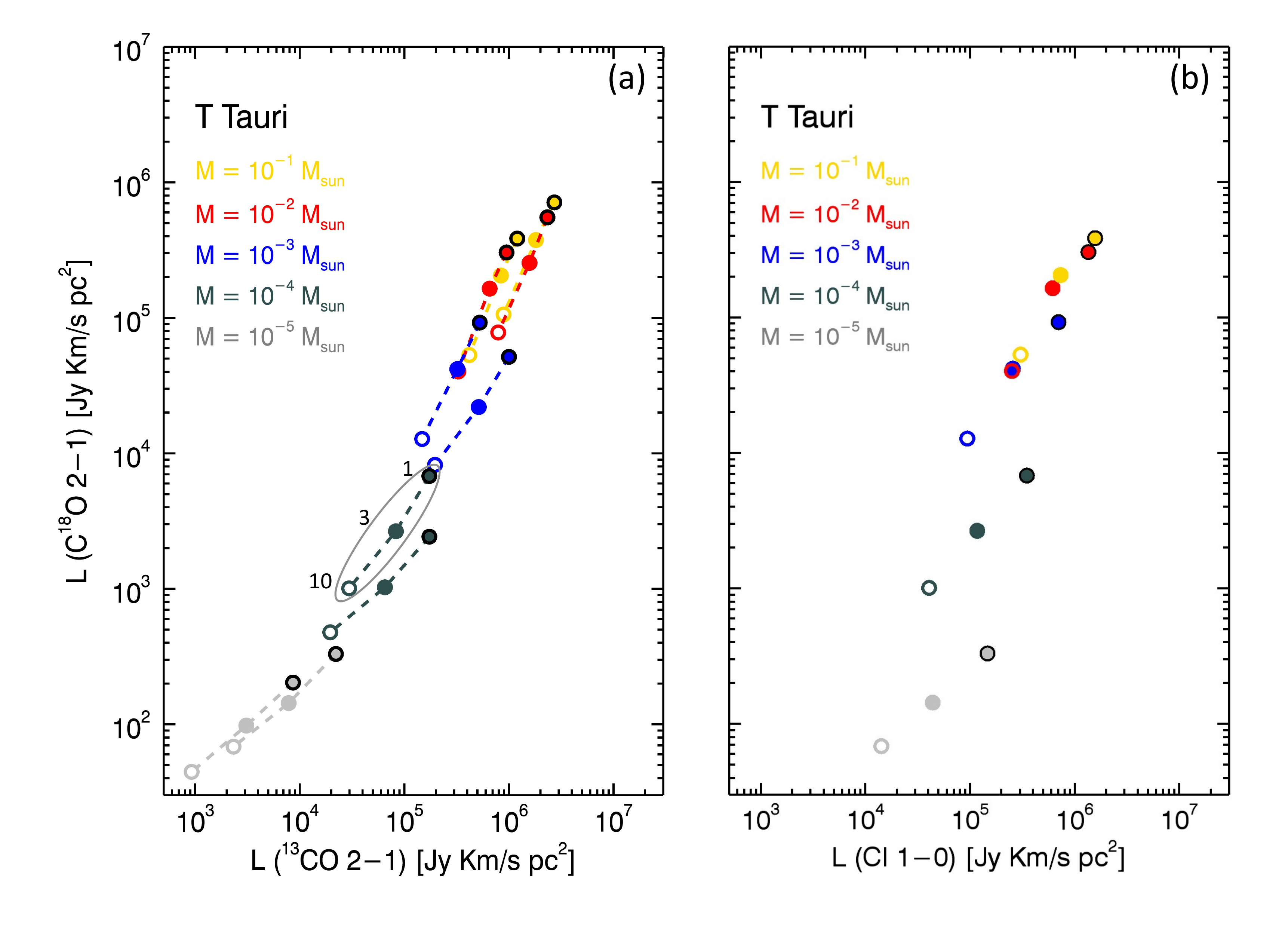}}
      \caption{ C$^{18}$O vs $^{13}$CO ($J=2-1$) (panel a) and C$^{18}$O vs [\ion{C}{i}] ($J=1-0$) (panel b) line luminosity for ten models, run also with a reduced initial atomic C and  $^{13}$C abundances. The standard value,  $X_{\rm C}=\text{[C]/[H]}=1.35 \times 10^{-4} $, has been reduced by factors of 3 and 10. The models with ISM-like carbon abundances are shown by the outlined symbols, those with a factor of 3 of carbon depletion by the filled symbols and those with a factor of 10 depletion by the unfilled symbols.}
       \label{C_depl}
\end{figure*}

Figures \ref{C_depl_ratios} and \ref{C_depl} show the resulting
C$^{18}$O and [\ion{C}{i}] line intensities for various carbon
abundances.  It is seen, also from Table \ref{C_red_fact}, that
C$^{18}$O line intensities scale sub-linearly with the level of carbon
depletion, i.e., the relative intensity decrease is smaller than the
relative elemental abundance change. This reflects the fact that the
C$^{18}$O abundances scale sub-linearly with the level of carbon
depletion in the outer disk, and super-linearly
in other regions. This in turn is due to the competing effects of
self-shielding and dust opacity, which can be illustrated as follows.

\begin{figure*}
\centering
             {\includegraphics[width=0.8\textwidth]{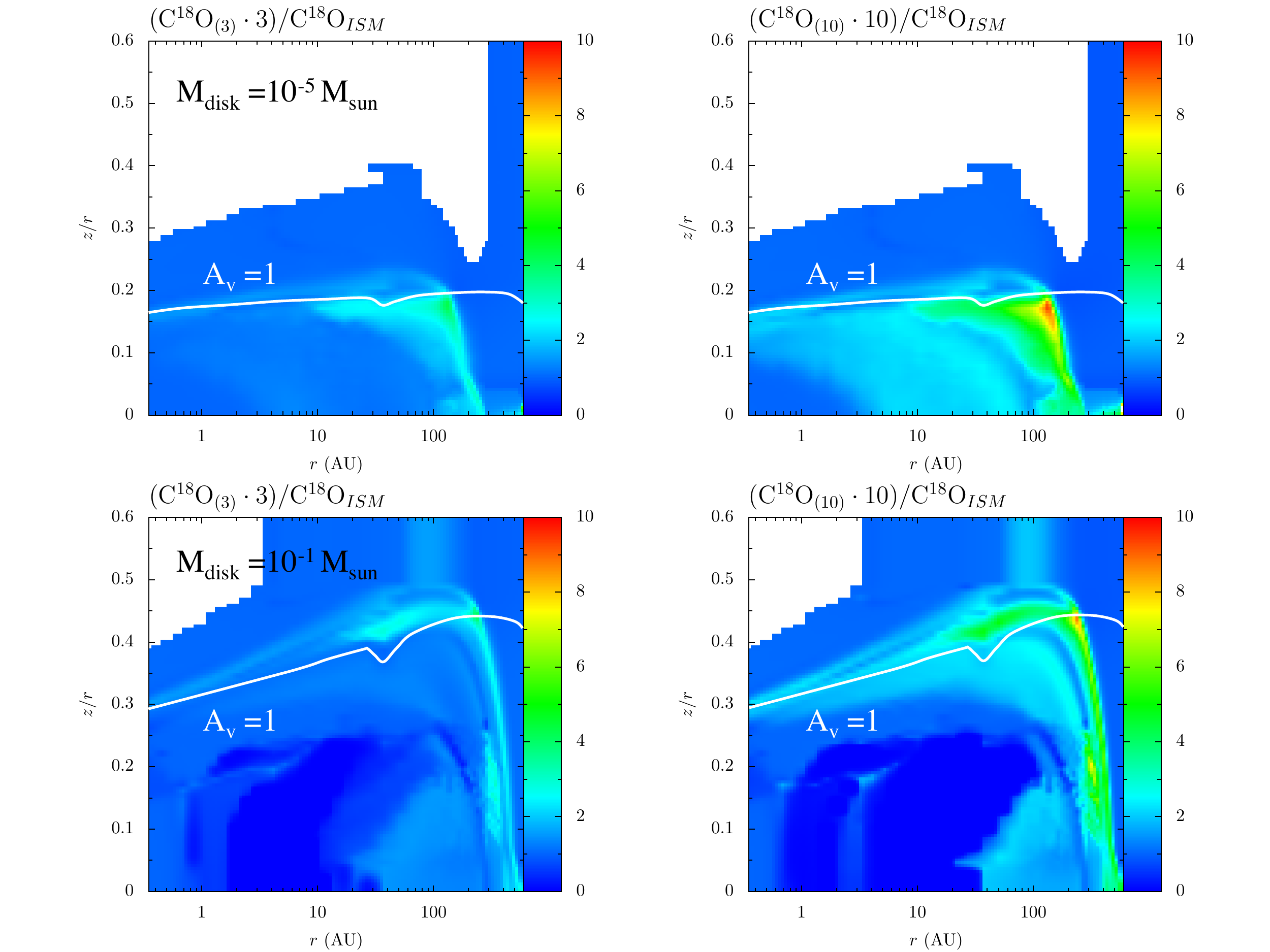}}
      \caption{Ratios of C$^{18}$O abundance obtained with a depleted level of carbon and with an ISM-like C abundance, rescaled by the level of carbon depletion. Where the ratio is higher than unity (colors from light blue to red), the C$^{18}$O abundance changes more slowly than the elemental carbon abundance. Upper panels show a $10^{-5}M_{\odot}$ T Tauri disk model, bottom panels show a $10^{-1}M_{\odot}$ T Tauri disk model. Other disk parameters are: $R_{\rm c}=60$ au, $\gamma$=0.8, $h_{\rm c}$=0.1, and $\psi$=0.1. The white solid line describes the A$_{\rm V}=1$ surface, where photodissociating photons are mainly absorbed by the dust.}
       \label{ratio_carbon}
\end{figure*}

Fig. \ref{ratio_carbon} shows the ratios of C$^{18}$O abundance
obtained with a depleted level of carbon and with an ISM-like C
abundance, rescaled by the level of carbon depletion. If the C$^{18}$O
abundance scaled linearly with carbon depletion, the ratio should be
equal to unity everywhere (color: medium blue). Where the ratio is
higher than unity (colors: from light blue to red), the C$^{18}$O
abundance changes more slowly than the elemental carbon
abundance. These regions are just below the $A_{\rm V}=1$ surface. The
carbon depletion decreases the C$^{18}$O abundance such that the dust
opacity begins to protect C$^{18}$O from photodissociation before
self-shielding becomes important. These regions contribute
significantly to the $J=2-1$ lines, while the $J=6-5$ originates
closer to the star, where the described effect is less important. As a
consequence the line luminosity ratios of C$^{18}$O $J=2-1$ over
$J=6-5$ transitions vary for different levels of carbon depletion, as
shown in panel (d) of Fig. \ref{C_depl_ratios}. Models with a ISM-like
carbon abundance have lower 2-1/6-5 line luminosity ratios than those
with lower carbon abundance.

On the other hand, it is seen in Fig. \ref{ratio_carbon} that there
are regions where the C$^{18}$O abundance changes faster than the
elemental carbon abundance (color: dark blue). These regions
contribute significantly to the $J=1-0$ lines. As a consequence the
line luminosity ratios of C$^{18}$O $J=1-0$ over $J=6-5$ transitions
vary for different levels of carbon depletion, as shown in panel (b)
of Fig. \ref{C_depl_ratios}. Models with a ISM-like carbon abundance
have higher 1-0/6-5 line luminosity ratios than those with lower
carbon abundance.  In both cases (sub-linear and super-linear
dependence on the carbon depletion) the effect on the line ratio is
larger for higher mass disks ($M_{\rm disk}=10^{-1}M_{\odot}$).

\end{appendix}
\end{document}